\documentclass[journal]{IEEEtran}

\usepackage{graphicx}
\usepackage{color}
\usepackage{epstopdf}
\usepackage{amsmath}
\usepackage{amsbsy}
\usepackage{amssymb}
\usepackage{amsthm}
\usepackage{subcaption}
\usepackage{epsfig}
\usepackage{psfrag}
\usepackage{algorithm}
\usepackage{algorithmic}
\usepackage[dvipsnames]{xcolor}
\usepackage{lipsum}
\usepackage{cancel}
\usepackage{tikz}
\usepackage{booktabs}
\usepackage{pifont}
\usepackage{mathtools}
\usepackage{flushend}
\usepackage{mathdots}
\usepackage{setspace}
\usepackage{microtype}
\usepackage{dsfont}

\usepackage[normalem]{ulem}

\newtheorem{proposition}{Proposition}

\newtheorem{definition}{Definition}

\newcommand{\cf}{\textrm{cf.}\ }
\newcommand{\eg}{\textit{e.g.}}
\newcommand{\ie}{\textit{i.e.}}

\newcommand{\via}{\textit{via }}
\newcommand{\viz}{\textit{viz.}}

\newcommand{\etc}{\textit{etc}}

\newcommand{\vide}{\textit{vide}\ }

\newcommand{\ba}{\boldsymbol{a}}

\newcommand{\be}{\boldsymbol{e}}

\newcommand{\bw}{\boldsymbol{w}}
\newcommand{\bx}{\boldsymbol{u}}
\newcommand{\by}{\boldsymbol{v}}
\newcommand{\bu}{\boldsymbol{u}}
\newcommand{\bv}{\boldsymbol{v}}
\newcommand{\bphi}{\boldsymbol{\varphi}}

\newcommand{\bE}{\mathbf{E}}
\newcommand{\bW}{\boldsymbol{W}}
\newcommand{\bX}{\boldsymbol{U}}
\newcommand{\bY}{\boldsymbol{V}}
\newcommand{\bA}{\boldsymbol{A}}
\newcommand{\bB}{\boldsymbol{B}}
\newcommand{\bU}{\boldsymbol{U}}
\newcommand{\bV}{\boldsymbol{V}}

\newcommand{\complex}[1]{\mathds{C}^{#1}}

\renewcommand{\natural}[1]{\mathds{N}^{#1}}
\newcommand{\calW}{\mathcal{W}}
\newcommand{\calU}{\mathcal{U}}

\newcommand{\T}{\textnormal{\textsf{T}}}
\renewcommand{\H}{\textnormal{\textsf{H}}}

\newcommand{\re}{\mathrm{Re}}

\newcommand{\tr}{\mathrm{tr}}
\newcommand{\brc}[1]{\left( #1 \right)}
\newcommand{\sqbrc}[1]{\left[ #1 \right]}
\newcommand{\figbrc}[1]{\left\{ #1 \right\} }

\newcommand{\abs}[1]{\left|#1\right|}
\newcommand{\norm}[1]{\left\|#1\right\|}
\newcommand{\e}{\mathrm{e}}
\renewcommand{\j}{\mathrm{j}}
\newcommand{\const}{\mathrm{const}}

\renewcommand{\vec}{\mathrm{vec}}
\newcommand{\maximize}{\mathrm{maximize}}
\newcommand{\st}{\mathrm{subject\;to}}

\renewcommand{\arg}{\mathrm{arg}}
\newcommand{\Mod}{\mathrm{\;mod\;}}

\newcommand{\dft}{\mathrm{DFT}}

\newcommand{\p}{\mathrm{P}}
\newcommand{\C}{\mathrm{C}}

\newcommand{\pt}{\mathrm{PT}}
\newcommand{\at}{\mathrm{AT}}

\begin{document}
\IEEEoverridecommandlockouts
    
\title{Efficient Cell-Specific Beamforming \\ for Large Antenna Arrays}

\author{Maksym~A.~Girnyk,~\IEEEmembership{Member,~IEEE,} and Sven~O.~Petersson
\thanks{Copyright \copyright~2021 IEEE. Personal use of this material is permitted. Permission from IEEE must be obtained for all other uses, in any current or future media, including reprinting/republishing this material for advertising or promotional purposes, creating new collective works, for resale or redistribution to servers or lists, or reuse of any copyrighted component of this work in other works.}
\thanks{%
M.~A.~Girnyk is with Ericsson Research, Stockholm, Sweden (e-mail: maksym.girnyk@ericsson.com). S.~O.~Petersson is with Ericsson Research, Gothenburg, Sweden (e-mail: sven.petersson@ericsson.com).}}

\markboth{To appear in IEEE Transactions on Communications}{Accepted paper}

\maketitle

\begin{abstract}
    We propose an efficient method for designing broad beams with spatially flat array factor and efficient power utilization for cell-specific coverage in communication systems equipped with large antenna arrays. To ensure full power efficiency, the method is restricted to phase-only weight manipulations. Our framework is based on the discovered connection between dual-polarized beamforming and polyphase Golay sequences. Exploiting this connection, we propose several methods for array expansion from smaller to larger sizes, while preserving the radiation pattern. In addition, to fill the gaps in the feasible array sizes, we introduce the concept of $\epsilon$-complementarity that relaxes the requirement on zero side lobes of the sum aperiodic autocorrelation function of a sequence pair. Furthermore, we develop a modified Great Deluge algorithm (MGDA) that finds $\epsilon$-complementary pairs of arbitrary length, and hence enables broad beamforming for arbitrarily-sized uniform linear arrays. We also discuss the extension of this approach to two-dimensional uniform rectangular arrays. Our numerical results demonstrate the superiority of the proposed approach with respect to existing beam-broadening methods.
\end{abstract}


\section{Introduction}
\label{sec:intro}
In spite of an entire list of new use cases---such as vehicle-to-vehicle, massive and critical machine-type communications---next-generation communication networks will remain being driven primarily by mobile broadband, demanding higher data rates. The corresponding radio interface, known as \emph{New Radio} (NR), is currently being developed by the 3GPP forum~\cite{3gpp2015study}. The technology is planned to be deployed in the mm-wave bands to relax the requirements on backwards compatibility with the legacy LTE networks~\cite[Ch.~24]{dahlman20164g}. After its initial successful deployment, NR is planned to gradually migrate to lower frequency bands and substitute LTE. 

Due to the specifics of the mm-wave channels, which exhibit clustering into few strong propagation paths, operation in the mm-wave spectrum is coupled with significant propagation losses~\cite{3gpp2018study}. To circumvent the subsequent coverage problems, beamforming techniques can be utilized. Small wavelengths in the mm-wave band enable very large antenna arrays that allow for large beamforming gains, albeit being of practically feasible physical sizes. These gains can be used to successfully compensate for the extreme path and penetration losses at high frequencies. 
Nonetheless, provided that every antenna element has a dedicated radio-frequency chain, large-array architectures become impractical due to increased cost and energy consumption. Hence, hybrid beamforming has recently become an attractive solution preserving the gains of large arrays, while reducing the number of digital chains~\cite{molisch2017hybrid}. The technique combines beamforming in both digital baseband and analog domain using phase shifters.

\begin{figure}
	\centering
	\begin{subfigure}{0.50\linewidth}
		\centering
		\includegraphics[width=4cm, trim=8.5cm 1.1cm 0cm 0cm, clip=true]{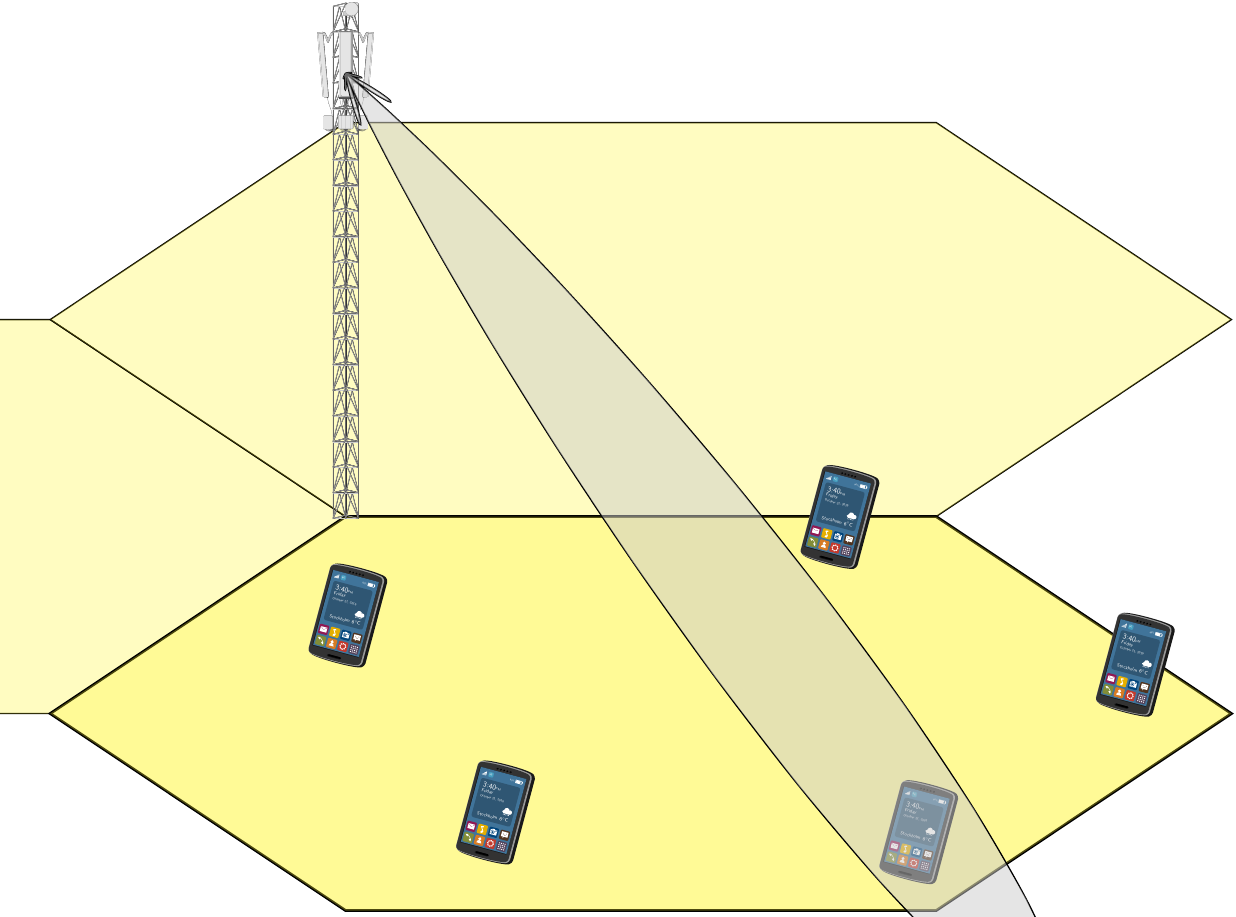}
		\caption{User-specific beamforming.}
		\label{fig:bfUeSpecific}
	\end{subfigure}%
	\begin{subfigure}{0.50\linewidth}
		\centering
		\includegraphics[width=4cm, trim=8.5cm 0cm 0cm 0cm, clip=true]{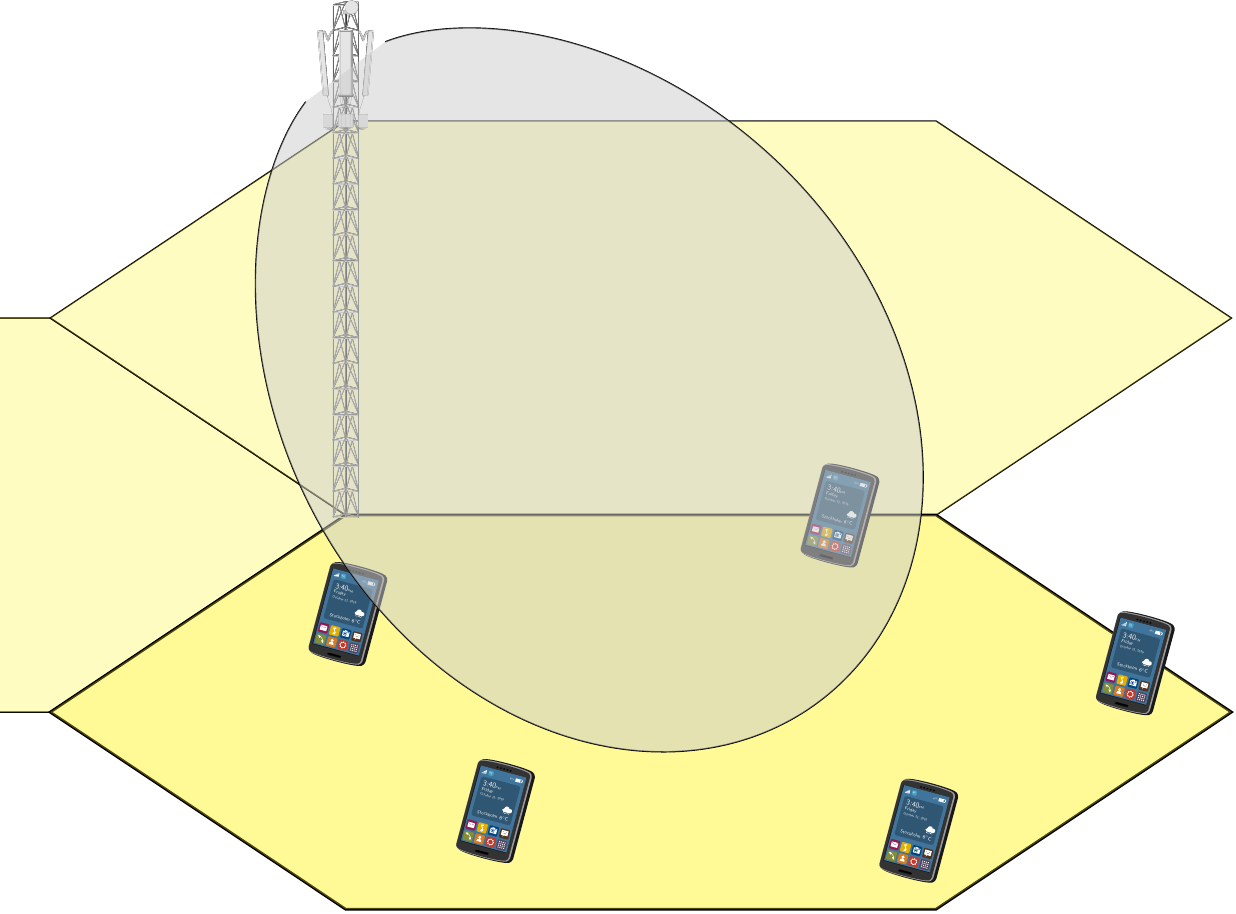}
		\caption{Cell-specific beamforming.}
		\label{fig:bfSectorSpecific}
	\end{subfigure}
	\vspace{0.25cm}
	\caption{Different types of beamforming.}
	\label{fig:beamforming}
	\vspace{-0.5cm}
\end{figure}

User-specific beamforming is known to be an efficient means for increasing the signal-to-noise ratio (SNR) at the receiver of a particular piece of user equipment (UE) by focusing the radiated energy towards the latter (see Fig.~\ref{fig:bfUeSpecific} for an illustration). A common approach to beamforming involves utilization of excitation weights in a form of a \emph{discrete-Fourier-transform} (DFT) vector which exhibits a number of beneficial features, such as maximum array gain, constant-modulus excitation weights and suitability for constructing precoding codebooks~\cite{love2003equal}. Meanwhile, in addition to the ability to focus energy at a particular UE with a narrow beam, it is also vital for a base station (BS) to be capable of forming broader beams to cover the entire cell with a desired level of radiation (see Fig.~\ref{fig:bfSectorSpecific}). In LTE and NR, this might be particularly relevant for control and broadcast channels, such as Physical Downlink Control Channel (PDCCH) and Physical Broadcast Channel (PBCH), as well as cell-specific signals, such as Primary Synchronization Signal (PSS), Secondary Synchronization Signal (SSS) or Cell-specific Reference Signal (CRS). Apart from synchronization and initial access, broad beams might be of use for mobility, small-packet and low-latency transmission~\cite{ericsson2017forming}. 


A \emph{broad beam}, defined herein as an array's radiation pattern whose \emph{half-power beamwidth} (HPBW) is equal to that of the pattern of a single antenna element, seems a natural solution for efficient cell-specific transmission of broadcast and control information. In practice, typical values of element HPBW range from $60^{\circ}$ to $90^{\circ}$, depending on the azimuth and elevation characteristics of the employed antenna elements. The design of broad beams using an entire antenna array is, however, rather problematic since direct enlargement of the array aperture leads to immediate beam shrinkage. Namely, even if individual antenna elements have an omnidirectional radiation pattern, the resulting pattern of the entire array gets narrower with increasing the number of elements. For example, an array with eight uniformly-radiating antenna elements, uniformly spaced half-wavelength apart and excited with the aforementioned DFT weights, produces a beam with HPBW of roughly $12.5^{\circ}$ which might be too narrow for cell-specific signaling. A typical three-sector deployment gives the angular sector width of $120^{\circ}$ which rather accommodates a set of roughly 10 such beams. Moreover, practical antenna systems employ a single power amplifier (PA) per antenna. Therefore, to efficiently utilize the available power resources, per-antenna power constraint should be satisfied. Hence, this paper aims at designing \emph{power-efficient broad beam} as a broad beam that is produced by constant-modulus beamforming weights, thereby utilizing all the available power at each antenna.

Various beamshape optimization algorithms~\cite{qiao2015broadbeam},~\cite{su2017semidefinite},~\cite{zhang2017energy} have been proposed to obtain broader beams. The downside with these approaches, though, is that variation in the amplitudes of the antenna excitation weights results in significant reduction in the output power due to poor utilization of the available PAs. 
An alternative approach---phase-only beamforming~\cite{brown2006extreme},~\cite{leifer2016revisiting},~\cite{intel2016codebook}---enables partial broadening of the beam without introducing any variation in the weight amplitudes. However, as proven in~\cite{qiao2015broadbeam}, it is not possible to obtain a broad beam without the latter. Moreover, a perfectly broad beam is \emph{only} achievable when a single antenna element (out of the entire array) is activated. 
Another method for broadcasting public information relies on performing a beam sweep over a set of available beams, also known as the grid-of-beams concept~\cite{saur2008grid}. The number of beams to consider can furthermore be reduced by exploiting orthogonal basis functions~\cite{thomas2015broadcast}. Other approaches to common-signal broadcasting include random beamforming~\cite{yang2012random}, space-time block coding~\cite{meng2016omnidirectional, xia2016space} and multi-stream precoding schemes~\cite{guo2018broad, lu2020omnidirectional, su2019omnidirectional} that exhibit omnidirectional sum power pattern over multiple resource blocks or streams. Nevertheless, these approaches typically do not guarantee a broad beam for single-layer transmission. 

To tackle the latter problem, a technique called array-size-invariant (ASI) beamforming was recently introduced in~\cite{petersson2020efficient}. The proposed approach is based on the fact that modern antenna systems are naturally built to exploit a pair of orthogonal polarizations. This dual-polarized nature provides an additional degree of freedom for designing a radiation pattern purely by means of phase-only techniques, removing the limitations reported in~\cite{qiao2015broadbeam}. Utilizing this fact, the method of~\cite{petersson2020efficient} allows for achieving a range of beamwidths by arbitrarily large antenna arrays, while guaranteeing efficient PA utilization. Unfortunately, the proposed method is realizable only with arrays of specific sizes, namely arrays of size $M=2^{a}$ in each dimension, where $a \in \natural{}_0$ due to the successive doubling of array size lying in the core of the ASI method. 

The present paper extends the existing ASI technique to arbitrarily-sized arrays. This is done by drawing a connection between the ASI beamforming and the concept of a \emph{complementary sequence pair} introduced by Golay in~\cite{golay1951static, golay1961complementary}. The latter refers to a pair of sequences whose aperiodic autocorrelation functions (AACFs) add up to a Dirac delta. We show that the ASI excitation weights designed to produce broad radiation patterns for a uniform linear array are, in essence, a pair of complex-valued \emph{polyphase} complementary sequences, applied per polarization. Similarly, for a uniform rectangular array, to obtain a broad beam, its excitation weights must be selected as a polyphase complementary array pair. 

Although Golay sequences have been studied extensively over the years (see~\cite{parker2003golay, sun2014survey} and references therein), large gaps remain in the knowledge of the sequence lengths for which they exist (see~\cite{eliahou1990new, craigen2002complex} for available lengths for known binary and quaternary Golay pairs, respectively). Furthermore, not much is known about the sufficient conditions for a sequence pair to form a Golay complementary pair~\cite{fiedler2013small}. Despite a lot of advances, only little is known about the lengths of polyphase Golay sequences. Most of the latest results are obtained \via various numerical searches~\cite{holzmann1994computer, ghaderpour2013asymptotic}. Unfortunately, though, no complex Golay sequences have been found so far for many practically relevant sequence lengths. 

To address this gap, the present paper introduces the concept of $\epsilon$-complementarity which relaxes the Golay condition and allows for a certain level of side lobes of the sum AACF. This opens a door to efficient heuristics for finding weight vectors that provide ``practically broad'' beams with efficient power utilization. One such heuristic algorithm---\emph{modified Great Deluge algorithm} (MGDA)---for designing broad beams by means of phase-only beamforming, is developed and presented herein. The proposed method enables the design of beams with various HPBW, up to that of an individual element. Since the method uses phase-only optimization, it is equally applicable to digital, analog and hybrid beamforming architectures. 

This paper is organized as follows. Sec.~\ref{sec:systemConfig} presents the system model and necessary background. Sec.~\ref{sec:broadBeamDesign} formulates the problem and reviews the state of the art. Next, Sec.~\ref{sec:broadBeamsArbitraryUla} draws a connection between broad beamforming and complementary sequences. In Sec.~\ref{sec:epsComplSequences}, $\epsilon$-complementary sequences are introduced and an algorithm that enables broad phase-only beamforming for arbitrarily-sized linear arrays is presented. Then, in Sec.~\ref{sec:broadBeamsUpa}, we discuss the generalization of the solution to two-dimensional planar arrays. In Sec.~\ref{sec:numericalExamples}, we present numerical examples to validate and illustrate the findings. The conclusions are drawn in Sec.~\ref{sec:conclusions}, while the appendices present discussions on practical aspects of the broad-beam operation.

The following notation is used throughout the paper. Upper-case and lower-case boldface letters (\eg, $\bA$ and $\ba$) denote matrices and column vectors, respectively. Hereafter, $\brc{\cdot}^{\T}$ denotes transpose, $\brc{\cdot}^*$ denotes conjugate,  $\brc{\cdot}^{\H}$ denotes Hermitian transpose, and $\re\figbrc{\cdot}$ denotes the real part of the argument. Moreover, $\delta (\cdot)$ denotes the Kronecker delta. Matrix $\mathbf{E}_M$ denotes the $M\times M$ exchange matrix, containing ones on the anti-diagonal and zeros everywhere else. The Kronecker product of two matrices $\bA$ and $\bB$ is denoted as $\bA\otimes\bB$, while the Hadamard product is denoted as $\bA\odot\bB$. The trace of a matrix $\bA$ is denoted as $\tr\figbrc{\bA}$, while $\vec\brc{\bA}$ denotes the vectorization operation, stacking the columns of $\bA$ into a large vector. Finally, $\calU [a, b)$ denotes the uniform distribution within interval $[a, b)$.

\section{System Configuration}
\label{sec:systemConfig}

Consider a base station (BS) that serves a sector with downlink common-signal transmission. The BS is equipped with an antenna array for the purpose of increased capacity and coverage. Typical array configurations utilized in practice are a one-dimensional linear array and a (rectangular) planar array. The necessary background on dual-polarized beamforming for these two array configurations is presented in the remainder of the present section.

\subsection{Uniform Linear Array}
\label{sec:ula}
\begin{figure}
	\centering
	\subcaptionbox{Uniform linear array.\label{fig:configUla}}[0.5\linewidth]
	{\centering
		\includegraphics[width=0.85\linewidth, trim=0cm -0.7cm 0cm 0cm, clip=true]{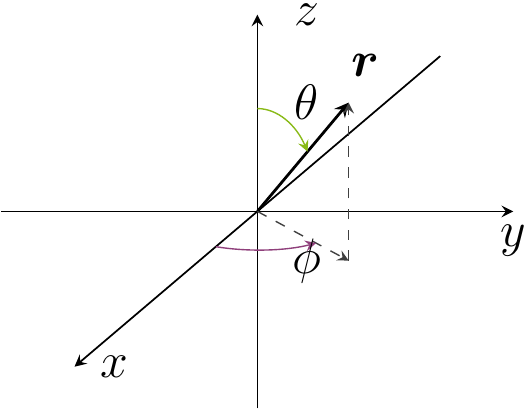}
		\includegraphics[width=0.85\linewidth, trim=0cm -0.9cm 0cm 0cm, clip=true]{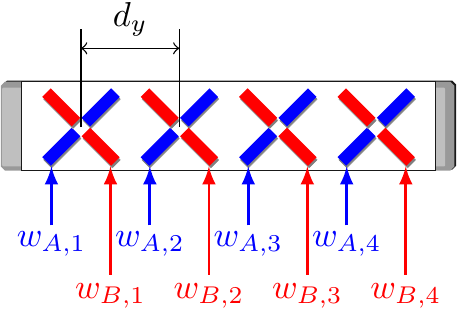}
		}%
	\subcaptionbox{Uniform rectangular array.\label{fig:configUpa}}[0.5\linewidth]
	{\centering
		\includegraphics[width=0.9\linewidth]{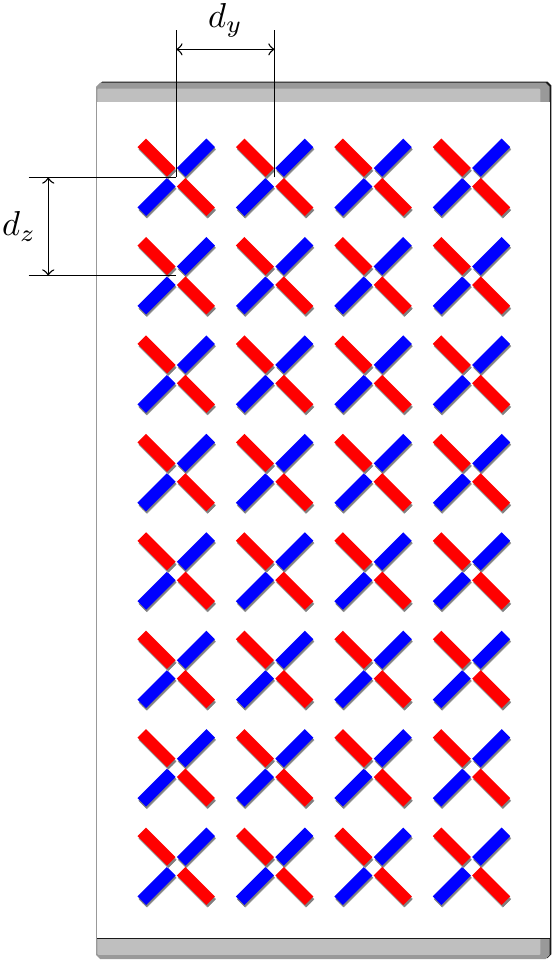}
		}
	\caption{One- and two-dimensional dual-polarized arrays.}
	\label{fig:configArray}
	\vspace{-0.5cm}
\end{figure}

Assume that a BS is equipped with an array of $M$ identical radiating elements, as depicted in Fig.~\ref{fig:configUla}. The antenna array has a line topology and the antenna elements are placed along the $y$-axis and spaced $d_{y}$ wavelengths apart from each other. Such an array configuration is referred to in the literature as a \emph{uniform linear array} (ULA)~\cite[Ch.~22]{orfanidis2002electromagnetic}. The individual elements adopted in typical modern antenna systems are dual-polarized---or cross-pole---antennas~\cite[Sec.~4.5.2]{asplund2020advanced}. Dual-polarized beamforming enables several benefits: co-locating more antennas at the same facility, utilization of polarization diversity schemes, as well as providing additional degrees of freedom for beam synthesis~\cite{petersson2020power}.

Each cross-pole element consists of two antenna ports with identical radiation patterns\footnote{Note that this assumption holds for an operational range of angles, not for all possible observation directions (see Appx.~\ref{sec:realisticElements}).} and orthogonal polarizations (\eg, vertical-horizontal or slanted $\pm 45^{\circ}$), referred to as polarizations $A$ and $B$ hereafter (\vide~Fig.~\ref{fig:configUla}). Thus, virtually, there are $2M$ antenna ports, and the ULA is excited with a pair of per-polarization vectors, yielding separate radiation patterns in each polarization. The polarization properties of the underlying antenna ports are modeled according to the angle-invariant model (also known as the 3GPP Model-2)~\cite[Sec.~7.1.1]{3gpp2017study}.\footnote{Note that a cross-pole under the angle-invariant model should not be confused with two crossed dipoles; such a depiction is merely used for the convenience of illustration. See Appx.~\ref{sec:realisticElements} for more details.} 

Following the IEEE standard~\cite{ieee2014standard}, we define the radiation pattern as follows.
\begin{definition}[Radiation power pattern]
    The total radiation power pattern of an antenna array is referred to as the spatial distribution of the power of the total electric field generated by the antenna array. Furthermore, the partial radiation power pattern of an antenna array for a given polarization is referred to as the spatial distribution of the radiated electric-field power corresponding to that particular polarization.
\end{definition}

Consider a ULA with $M$ cross-pole antennas, depicted in Fig.~\ref{fig:configUla} alongside the adopted coordinate system. The antenna ports are excited with complex-valued weights $w_{A,m}$ and $w_{B,m}$ for $m\in\{1,\ldots,M\}$, at respective polarizations. Due to the practical per-antenna power constraint and with aim of efficient PA usage, all the available power per antenna should be utilized. Therefore, the excitation weights are restricted to have fixed modulus. Without loss of generality, we consider unimodular excitations, \ie, $|w_{A,m}| = |w_{B,m}| = 1,\;\forall m \in\figbrc{1,\ldots,M}$. Consider the electric field radiated by the ULA; in the rotated $\brc{A, B}$-basis it is given by a vector\footnote{It should be noted that for the sake of presentation simplicity, all the irrelevant terms, such as constants, dependencies on time and actual observation radius, have been omitted in the equations hereafter.} 
\begin{equation}
\label{eqn:eFieldsUla}
    \be\brc{\phi, \theta} = \sqbrc{\begin{matrix}
           	\bw_{A}^{\T} \ba\brc{\psi}\\
           	\bw_{B}^{\T} \ba\brc{\psi}
           	\end{matrix}} E_0\brc{\phi,\theta},
\end{equation}	
where $E_0\brc{\phi,\theta}$ is the electric field radiated by a single cross-pole, assumed identical for both polarizations, and $\ba(\psi)$ is the steering vector of the ULA given by a Vandermonde vector
\begin{equation}
\label{eqn:steeringVectorUla}
   	\ba(\psi) = [1,\; \e^{\j\psi(\phi,\theta)}, \ldots, \e^{\j(M-1)\psi(\phi,\theta)} ]^{\T} \in \complex{M},
\end{equation}
where, for a horizontal ULA (with cross-poles along the $y$-axis), the relative phase shift reads
\begin{equation}
\label{eqn:psiUla}
    \psi(\phi,\theta) \triangleq 2 \pi d_{y} \sin\theta \sin\phi.
\end{equation}
A receiver, thus, observes the superposition of the fields radiated at both polarizations. Because a typical UE employs two antennas operating on two orthogonal polarizations, the UE is able to pick all the energy across the polarizations. It is furthermore assumed that the UE antennas have equal gains and orthogonal polarizations in all directions.\footnote{Note that this clearly is an idealistic assumption. However, the effect of non-ideal receiver antennas would be a form of polarization mismatch, a phenomenon that also exists for conventional single-polarized beamforming with non-ideal receivers.}

To counteract performance losses due to unknown orientation of the UE and unknown, due to the impact of the radio channel, polarization at the UE antennas, diversity reception has become conventional for UE receivers. For instance, the maximum-ratio combining (MRC) is a widely used scheme which combines the powers from the different receiver antennas. This is equivalent to seeing at the receiver end a single beam whose polarization state is varying with the observation angle.

When applying the MRC operation, the receiver sees the total radiation power pattern
\begin{equation}
   	G\brc{\phi,\theta} = \be^{\H}\brc{\phi,\theta} \be\brc{\phi,\theta}
   	= \abs{A(\phi,\theta)}^2 \; G_0(\phi,\theta),
   	\label{eqn:patternMultiplication}
\end{equation}
$G_0(\phi,\theta) = \abs{E_0\brc{\phi,\theta}}^2$ is the element power pattern and $|A(\phi,\theta)|^2$ is the \emph{power-domain array factor} of the dual-polarized ULA.\footnote{Factorization~\eqref{eqn:patternMultiplication} represents the pattern-multiplication property~\cite[Ch.~20]{orfanidis2002electromagnetic} for a dual-polarized ULA in the power domain.} For a pair of orthogonal polarizations $(A, B)$ the latter can be expressed as
\begin{equation}
\label{eqn:arrayFactorDefinition}
   	|A(\phi,\theta)|^2 = \abs{\bw_A^{\T} \ba(\psi)}^2 + \abs{\bw_B^{\T} \ba(\psi)}^2.
\end{equation}
Note that the steering vector in~\eqref{eqn:steeringVectorUla} can be multiplied by an arbitrary constant-modulus complex number, which would correspond to a fixed phase shift or time delay of the transmission at each antenna of the array, having no impact on its power characteristics.

\subsection{Uniform Rectangular Array}
\label{sec:upa}
In the two-dimensional case, where the antenna elements are placed on a rectangular grid, the array is referred to as a \emph{uniform rectangular array} (URA)~\cite[Ch.~22]{orfanidis2002electromagnetic}. For example, the antenna array at the BS can be formed by $M$ columns with $N$ antennas each. The excitation weights in this scenario are formed by a pair of matrices $\brc{\bW_A, \bW_B}$ of corresponding sizes. When applied to a URA, the weights produce a three-dimensional radiation pattern, which is factorized similarly to that of a ULA~\eqref{eqn:arrayFactorDefinition}. Thus, the power-domain array factor reads
\begin{equation}
	|A\brc{\phi, \theta}|^2 = \abs{\tr\figbrc{\!\bW_A^{\T}\bA\brc{\psi_y, \psi_z}}}^2\!\!\!+\abs{\tr\figbrc{\!\bW_B^{\T}\bA\brc{\psi_y, \psi_z}}}^2\!\!\!,
\end{equation}
where $\bA\brc{\psi_y, \psi_z}$ is the steering matrix of the array given by 
\begin{equation}
	\bA\brc{\psi_y, \psi_z} = \ba_y^{\T}\brc{\psi_y} \otimes \ba_z\brc{\psi_z}  \in \complex{N\times M},
\end{equation}
that is, a combination of steering vectors for both dimensions
\begin{subequations}
	\begin{align}
	\ba_y(\psi_y) &= [1,\; \e^{\j\psi_y(\phi,\theta)}, \ldots, \e^{\j(M-1)\psi_y(\phi,\theta)} ]^{\T}\in \complex{M},\\
	\ba_z(\psi_z) &= [1,\; \e^{\j\psi_z(\phi,\theta)}, \ldots, \e^{\j(N-1)\psi_z(\phi,\theta)} ]^{\T}\in \complex{N}.
	\end{align}	
\end{subequations}
Here the relative phase shifts are defined as
\begin{equation}
    \psi_y(\phi,\theta) \triangleq 2 \pi d_{y} \sin\theta \sin\phi,\qquad
    \psi_z(\phi,\theta) \triangleq 2 \pi d_{z} \cos\theta.
\end{equation}

\section{Broad Beam Design}
\label{sec:broadBeamDesign}
This section discusses the problem of designing broad beams and reviews the state-of-the-art approaches. Before formulating the problem of broad-beam design, we first formally define the \emph{broad beam} concept as follows.
\begin{definition}[Broad beam]
\label{def:broadBeam}
    A broad beam is referred to as the total radiation pattern whose the power-domain array factor is spatially flat, \ie,
    \begin{equation}
    \label{eqn:defBroadBeam}
       \abs{A\brc{\phi, \theta}}^2 = \const, \quad \forall \phi,\theta.
    \end{equation}
\end{definition}
In addition to the above, the beams of interest also incorporate the power efficiency requirement. 
\begin{definition}[Power-efficient broad beam]
\label{def:broadBeamPe}
    A power-efficient broad beam is referred to as the broad beam according to Def.~\ref{def:broadBeam} produced by excitation weights $\bw_A$ and $\bw_B$ that are unimodular, \ie, $|w_{A,m}| = |w_{B,m}| = 1,\;\forall m \in\figbrc{1,\ldots,M}$.
\end{definition}
 
Considering the above, the main problem tackled in this paper can be formulated as follows: 
\begin{quote}
\emph{Find proper unimodular excitation weights whose total radiation pattern's HPBW is as broad as that of a single element's radiation pattern.}
\end{quote}
    
Unfortunately, it is not possible to obtain the desired weights by conventional beamforming methods. In fact, the width of the beam produced by an antenna array, using conventional beamforming methods, is inversely proportional to the aperture size of the array (unless amplitude taper is applied). Consider, for example, a conventional approach, utilized in both LTE and NR, based on exciting the ULA with a DFT vector, picked from a codebook of precoders (see, \eg,~\cite[Ch.~10]{dahlman20164g}). Such a beamformer is given by a linear phase front with certain pointing azimuth angle $\phi_0$ (for a horizontal array), \ie,
\begin{equation}
\label{eqn:weightsDft}
    \bw_{\dft} = \sqbrc{1, \e^{-\j\psi(\phi_0,\theta)}, \ldots, \e^{-\j (M-1) \psi(\phi_0,\theta)}}^{\T} \in \complex{M}
\end{equation}
at both orthogonal polarizations. The shape of such a DFT beam is determined by the parameters of the ULA and is described by a Dirichlet kernel~\cite[Ch.~2]{davis1989fourier}  
\begin{equation}
    D\brc{\phi, \theta} = \frac{\sin\sqbrc{\frac{M}{2}\pi d_{y} \sin \theta \brc{\sin \phi - \sin \phi_0}}}{\sin\sqbrc{\frac{1}{2}\pi d_{y} \sin\theta \brc{\sin \phi - \sin \phi_0}}}.
\end{equation}
As the number or antennas, $M$, grows large, and provided that $\abs{\phi-\phi_0} < \pi$, the Dirichlet kernel above tends to a Dirac delta function in $\phi$. That is, the HPBW of a DFT beam inevitably reduces as the size of the ULA increases, focusing the radiated energy in a given angular direction.

\subsection{Existing Approaches to Broad Beamforming}
Even though the above beamforming method can be beneficial for the UE-specific transmission in a cellular system, increasing the SNR of a given UE, in order to broadcast control information to all the UEs within a sector one needs to be able to form a broader beam. As an extreme, one could aim at obtaining a pattern of the same HPBW as that of a radiation pattern of an individual element. Below we review a few representative beamforming approaches to this task.

\subsubsection{Amplitude tapering}
One way to broaden the beam is to allow for the optimization of the amplitudes of the excitation weights. Unfortunately, any variation in the weight amplitudes leads to power loss due to the inefficient utilization of the available PAs. Moreover, in~\cite{qiao2015broadbeam}, it was proven that for a given polarization the only solution that guarantees angle-independent array factor is the transmission from a single element of the array. This solution is clearly impractical due to its wasteful utilization of the available power.

To overcome the above issue, the approach of~\cite{qiao2015broadbeam} allows for an angular fluctuation of the power-domain array factor, $\epsilon(\phi)$, while minimizing the peak-to-average-power ratio, \ie,
\begin{equation}
\label{eqn:weightsQiao}
    \bw_{\at} = \underset{\bw\in\calW}{\arg \min} \figbrc{\frac{M\max_m \abs{w_m}^2}{\norm{\bw}^2}},
\end{equation}
over a set of weights ensuring that the fluctuation of the array factor is bounded by $\epsilon(\phi)$, \viz,
\begin{equation}
\label{eqn:weightSpaceQiao}
    \calW = \figbrc{\bw\!: \abs{\bw^{\T} \ba(\phi)}^2 \!\!= M \sqbrc{1 + \epsilon(\phi)}, \forall \phi \in \! \sqbrc{-\frac{\pi}{2}, \frac{\pi}{2}}}.
\end{equation}
By choosing a bounded error function, \eg, $\epsilon(\phi) = \zeta \cos 2\phi$, where $\zeta$ is a tolerance level, and sweeping over all solutions to a polynomial equation, one obtains an omnidirectional beam with a small level of beam ripple.

Nevertheless, in spite of achieving a spatially flat array factor, the solution proposed in~\cite{qiao2015broadbeam} exhibits variations in the amplitudes of the excitation weights. As discussed before, this inevitably leads to poor PA utilization and subsequent power losses. For example, a loss of $8$ dB due to amplitude taper was reported in~\cite{qiao2015broadbeam} for a ULA with $M=128$ antennas. Hence, even though this beam is broad, it is not power-efficient according to Def.~\ref{def:broadBeamPe}.

\subsubsection{Phase tapering}
As an alternative to the above approach, one could turn to phase-only beamforming. For instance, a practical method for beam broadening was proposed in~\cite{intel2016codebook}. The main idea lies in applying a broadener function $f_m (p,c)$ on a base DFT weight vector~\eqref{eqn:weightsDft}. For an $M$-antenna ULA, the phase-tapering weights are chosen according to
\begin{equation}
\label{eqn:weightsIntel}
    \bw_{\pt} = \bw_{\dft}\odot[\e^{\j f_1\brc{p,c}}, \ldots, \e^{\j f_{M} \brc{p,c} } ]^{\T},
\end{equation}
where the broadener function is given by
\begin{equation}
    f_m (p,c) = \abs{4\pi c\brc{\frac{2m-M-1}{2(M-1)}}^p },
\end{equation}
for all $m\in\figbrc{1,\ldots,M}$. Parameters $(p,c)$ allow for further optimization of the beam shape given an immediate objective. 
Although the above solution achieves certain level of beam broadening, we recall that~\cite{qiao2015broadbeam} has proven that for a given polarization it is \emph{impossible} to construct a broad beam by means of phase-only tapering. Indeed, even though this beam is power-efficient, it is not broad according to Def.~\ref{def:broadBeam}.

\subsubsection{Array-size invariant beamforming}
To circumvent the shortcomings of the above solutions, the ASI beamforming approach was proposed in~\cite{petersson2020efficient}, providing a means of creating beams of controllable width. The basic underlying principle lies in the exploitation of the additional degrees of freedom coming from the utilization of polarization inherent to the dual-polarized beamforming~\cite{petersson2020power}. Namely, a dual-polarized antenna array at the transmitter allows the creation of a total power radiation pattern whose polarization state varies with the observation angle. Two orthogonally polarized antennas at the receiver can then pick up all the power of the total beam across both the polarizations. Remarkably, the aforementioned extra degrees of freedom can be utilized for efficient design of various beam shapes~\cite{petersson2020power}. 

The main technique of~\cite{petersson2020efficient}, referred to as \emph{array expansion}, is based on successive doubling of the size of an array, while preserving its radiation pattern. Suppose that a dual-polarized ULA with $M$ antennas---referred to as a \emph{protoarray}---has been designed to have some desired total radiation pattern. The beamforming weights of the protoarray are given by two per-polarization vectors $\bw_{\p, A}$ and $\bw_{\p, B}$. The main task of the array expansion is to design a \emph{companion array} (with beamforming weights $\bw_{\C, A}$ and $\bw_{\C, B}$), which is appended to the protoarray, to form an expanded array of size $2M$ preserving the shape of the protoarray's pattern. The companion array shall, thus, have the same topology and same type of elements as the protoarray.

According to~\cite{petersson2020efficient}, an expanded array with weights
\begin{equation}
\label{eqn:weightsExpandedArray1d}
  	\bw_A = [\bw_{\C, A}^{\T}, \; \bw_{\p, A}^{\T}]^{\T},\qquad
  	\bw_B = [\bw_{\C, B}^{\T},\; \bw_{\p, B}^{\T}]^{\T},
\end{equation}
preserves the radiation pattern of the protoarray, provided that the companion array is excited~with
\begin{equation}
\label{eqn:weightsCompanionArray1d}
  	\bw_{\C, A} = - \bE_M \bw_{\p, B}^*,\qquad
  	\bw_{\C, B} = \phantom{-} \bE_M \bw_{\p, A}^*.
\end{equation}
The choice of weights for the companion array in~\eqref{eqn:weightsCompanionArray1d} leads to radiated electric fields of the subarrays being orthogonal regardless of the observation angle. In this way, the beams add up \emph{incoherently} and the effective beam does not get narrower. 

If the radiation pattern of the protoarray is broad and power efficient, according to Def.~\ref{def:broadBeamPe}, so is the pattern of the expanded array. Furthermore, the above expansion procedure can be applied recursively $k$ times (\ie, $k$-fold expansion), increasing the size of the expanded array to $2^k M$.

It is also noteworthy that the procedure is directly related to the Alamouti beamforming~\cite{girnyk2020simple}. Namely, it assures that the dual-polarized elements of the protoarray and its companion array are pair-wise conjugated in the Alamouti fashion:

\begin{equation}
	w_{\C,A,m} \!=\! - w^*_{\p,B,M-m+1},\quad
	w_{\C,B,m} \!=\!  w^*_{\p,A,M-m+1}.
\end{equation}
In digital communications, this principle is known as time-reversal Alamouti space-time coding, and it has been utilized in multi-carrier communication systems~\cite{vook2000transmit},~\cite{lindskog2000transmit}.

\section{Power-Efficient Broad Beams for Arbitrary-Sized Arrays}
\label{sec:broadBeamsArbitraryUla}

The aforementioned ASI beamforming~\cite{petersson2020efficient} relies on successive doubling of the ULA, thereby providing means to design pattern-preserving weights for ULAs with sizes $M=2^a D$, where $a \in \natural{}_0$ and $D$ is the size of the protoarray. Practical arrays, however, oftentimes feature other sizes, and hence there is a need to generalize this method to accommodate all the moderate integer sizes. The remainder of the section aims at filling this gap by providing a solution to the problem of broad-beam design for ULAs of arbitrary size. 

\subsection{Golay Sequences}
We start with identifying the similarity between the desired excitation-weight vectors and the so-called \emph{Golay complementary sequence pairs} that were introduced by Golay in his seminal work~\cite{golay1951static} related to infrared spectrometry. To describe the concept of Golay sequences, we first need to define the following function.

\begin{definition}[AACF]
\label{def:defAacf}
    Let $\bu \triangleq [u_1, u_2, \ldots, u_M]^{\T} \in \complex{M}$ be a sequence/vector of complex unit-norm values. Its aperiodic autocorrelation function (AACF) is defined as
    \begin{align}
    \label{eqn:defAacf}
    	R_{\bu}(\tau) = \left\{ 
    	\begin{matrix} 
    	&\sum\limits_{m=1}^{M-\tau} u_m u_{m+\tau}^*, 
    	& & 0\leq \tau \leq M-1,\\
    	&\sum\limits_{m=1}^{M+\tau} u_{m-\tau} u_{m}^*, 
    	& &-M+1\leq \tau < 0,\\
    	& 0,
    	& & \tau \notin (-M,M).
    	\end{matrix} \right. 
    \end{align}
\end{definition}
Based on the above definition, complex-valued Golay complementary sequences ~\cite{sivaswamy1978multiphase},~\cite{frank1980polyphase} are defined as follows.
\begin{definition}[Golay sequence pair]
\label{def:defGolayPair}
    A pair of unimodular sequences $\brc{\bu, \bv}$ is said to form a Golay complementary pair if it holds that
    \begin{align}
    \label{eqn:conditionGolay}
    	R_{\bu}(\tau) + R_{\bv}(\tau) &= 2M \delta (\tau).
    \end{align}
    Furthermore, a sequence $\bu\in\complex{M}$ that forms a Golay pair with another sequence $\bv\in\complex{M}$ is called a Golay sequence.
\end{definition}

The original complementary sequences proposed in~\cite{golay1961complementary} were binary, \viz, $u_m \in \{\pm 1\}, \; \forall m$. These are often referred in the literature to as ``real'' Golay sequences. At the same time, the concept can be generalized to larger alphabets. For instance, when the possible values of an entry are restricted to $u_m\in\{\pm 1, \pm \j\}, \, \forall m$, the sequences are called quaternary, or sometimes ``complex'', Golay sequences~\cite{craigen2002complex}. This can be extended further, to \emph{polyphase} Golay sequences whose entries include all values on a unit circle~\cite{sivaswamy1978multiphase},~\cite{frank1980polyphase}. The sequences have also been generalized to, \eg, QAM Golay sequences~\cite{roessing2001construction}, ternary Golay sequences~\cite{gavish1994ternary}, $Z$-complementary sequences~\cite{fan2007z}, dihedral complementary sequences~\cite{kamali1998dihedral}, complementary arrays~\cite{jedwab2007golay} and sets~\cite{tseng1972complementary}. Golay sequences are also shown to have tight connection to Barker sequences~\cite{jedwab2009construction}, Hadamard matrices~\cite{turyn1974hadamard}, and second-order Reed-Muller codes~\cite{davis1999peak}. 

\subsection{Known Sequence Lengths}
Unfortunately, Golay sequences might exist not for all possible lengths---at least in the finite-alphabet case. The length of an existent Golay sequence is referred to as a \emph{Golay number}. Already in~\cite{golay1961complementary}, it was shown that for binary pairs any Golay number must constitute a sum of two squares. In addition, a method for constructing complementary sequences of length $2^a$, where $a\in \mathbb{N}_0$ was also provided already. Later, it was proven in~\cite{eliahou1990new} that no binary Golay number is divisible by a prime number congruent to $3\Mod4$. To date, it is known that binary Golay sequences exist for lengths $M = 2^a 10^b 26^c, \; \forall a,b,c \in \natural{}_0$~\cite{eliahou1990new}.

As was suggested in~\cite{frank1980polyphase}, the above restrictions disappear for quaternary Golay pairs. It was furthermore shown \via an exhaustive numerical search~\cite{holzmann1994computer, ghaderpour2013asymptotic} that quaternary Golay sequences do not exist for lengths $7, 9, 14, 15, 17, 19, 21$, \etc. To date, quaternary Golay sequences are only known for $M = 2^{a+f} 3^b 5^c 11^d 13^e$, where $f \leq c+e,\; b+c+d+e \leq a+2f+1,\; a,b,c,d,e,f \in \natural{}_0$, \cf~\cite[Cor.~4]{craigen2002complex}. Apart from the above, only little is known about the lengths for which polyphase Golay sequences exist~\cite{fiedler2013small}.

\subsection{Connection to ASI Beamforming}
The following proposition shows the connection between Golay sequences and dual-polarized beamforming~\cite{petersson2020power}.
\begin{proposition}
\label{prop:golayAsiEquivalence}	
    For a dual-polarized ULA, condition~\eqref{eqn:defBroadBeam} is fulfilled if the pair of per-polarization beamforming vectors $(\bw_A, \bw_B)$ is chosen as a Golay complementary sequence~pair.
\end{proposition}

\begin{IEEEproof}
    Let $S_{\bu}(f)$ be the power spectral density (PSD) of a function whose discrete values are given by sequence $\bu$. It is known that the PSD is connected to the AACF~\eqref{eqn:defAacf} through the Wiener-Khintchine transform~\cite[Ch. 5]{stranneby2004digital}
    \begin{align}
        S_{\bu}(f) &= \sum_{\tau=-M+1}^{M-1} R_{\bu}(\tau) \; \e^{-\j 2\pi f \tau}.
    \end{align}
    Hence, due to the particular form of the sum-AACF of a Golay complementary pair of sequences $(\bu,\bv)$, implies that their power spectra add up to a constant, \ie,
    \begin{align}
    \label{eqn:golayPairSpectra}
        S_{\bu}(f) + S_{\bv}(f) = 2M.
    \end{align}
    
    Since the per-polarization radiation field patterns of a ULA, excited with weights $\bw_A$ and $\bw_B$, are spatial Fourier transforms of the weights sequences, we have 
    \begin{align}
    \label{eqn:asiWeightsSpectra}
        S_{\bw_A}(\psi) + S_{\bw_B}(\psi) = 2M.
    \end{align}
    Therefore, condition~\eqref{eqn:defBroadBeam} is equivalent to dual-polarized beamforming with a pair of excitation vectors $(\bw_A, \bw_B)$ being a (polyphase) Golay complementary pair. 
\end{IEEEproof}

The above result shows that to obtain weights forming a power-efficient broad beam one needs to find a pair of complementary sequences of corresponding length. The following proposition provides a recursive construction for designing larger broad-beam weight vectors from smaller ones, thereby extending the set of known ASI beamforming vectors~\cite{petersson2020efficient}.

\begin{proposition}
\label{prp:golayConstructionUla}
	Assume that a protoarray of size $M$ excited with beamforming weights $(\bw_{\p,A}, \bw_{\p,B})$ produces a desired radiation pattern. Furthermore, let $(\bx, \by)$ be a polyphase Golay sequence pair of length $N$. Then the expanded array of size $2NM$ with excitation vectors
    \begin{align}
    	\label{eqn:golayConstructionUla}
    	\!\!\!\!\bw_A \!=\! \sqbrc{\begin{matrix}
    		\bx \otimes \bw_{\p,A}\\
    		-\by \otimes \brc{\bE_M\bw_{\p,B}^*}
    	\end{matrix}},\;\;\;
    	\bw_B \!=\! \sqbrc{\begin{matrix}
    		\bx \otimes \bw_{\p,B}\\
    		\by \otimes \brc{\bE_M\bw_{\p,A}^*}
    		\end{matrix}}\!,
    \end{align}
    preserves the radiation pattern of the protoarray.
\end{proposition}

\begin{IEEEproof}
Consider the antenna array obtained in~\eqref{eqn:golayConstructionUla}. It consists of two parts, each excited with a pair of per-polarization weight vectors given by
\begin{align}
\label{eqn:golaySubarraysUla}
	\bw_{1, A} &= \phantom{-}\bx \otimes \bw_{\p,A},
	& & \bw_{1, B} = \bx \otimes \bw_{\p,B},\\
	\bw_{2, A} &= -\by \otimes \bE_M\bw_{\p,B}^*, 
	& & \bw_{2, B} = \by \otimes \bE_M\bw_{\p,A}^*.
\end{align}
The electric-field vector of the expanded array is the superposition of the fields of its two parts
\begin{align}
    \!\!\be\brc{\phi, \theta} 
    &=\be_{1}\brc{\phi, \theta} + \be_{2}\brc{\phi, \theta}\\
    &=\brc{\sqbrc{\begin{matrix}
    	\bw_{1, A}^{\T} \ba_{1}(\psi)\\
    	\bw_{1, B}^{\T} \ba_{1}(\psi)
    	\end{matrix}} \!+\!
    	\sqbrc{\begin{matrix}
    	\bw_{2, A}^{\T} \ba_{2}(\psi)\\
    	\bw_{2, B}^{\T} \ba_{2}(\psi)
    	\end{matrix}} } E_0\brc{\phi,\theta},
\end{align}
where the steering vector for the expanded array is given by
\begin{align}
    \ba(\psi) & = [\ba_{1}(\psi)^{\T}, \; \ba_{2}(\psi)^{\T}]^{\T} \\
    & = [1,\; \e^{\j\psi(\phi,\theta)}, \ldots, \e^{\j2(NM-1)\psi(\phi,\theta)} ]^{\T} \in \complex{2NM},
\end{align}
with $\ba_{1}(\psi)$ and $\ba_{2}(\psi)$ being the steering vectors of the two subarrays. Performing the MRC operation at the receiver on the two field components, we get the total radiation pattern as
\begin{align}
	G\brc{\phi, \theta} 
	=&\;\be^{\H}\brc{\phi, \theta}\be\brc{\phi, \theta}\\
	=&\norm{\be_{1}\brc{\phi, \theta}}^2 + \norm{\be_{2}\brc{\phi, \theta}}^2 \nonumber\\
	&\qquad+ 2 \;\re\figbrc{\be_{1}^{\H}\brc{\phi, \theta} \be_{2}\brc{\phi, \theta}}.
\label{eqn:totalPatternGolayUla}
\end{align}

First, consider the inner product
\begin{align}
	\be_{1}^{\H}&\brc{\phi, \theta}  \be_{2}\brc{\phi, \theta}\nonumber \\ 
	&= \left[\ba_{1}^{\H}(\psi)\bw_{1, A}^{*} \bw_{2, A}^{\T} \ba_{2}(\psi) \right. \nonumber \\
	&\quad+ \left.\ba_{1}^{\H}(\psi)\bw_{1, B}^{*} \bw_{2, B}^{\T} \ba_{2}(\psi) \right] G_0\brc{\phi,\theta} \\
	&= \left[-\ba_{1}^{\H}(\psi) \brc{\bx^{*} \by^{\T}} \otimes \brc{\bw_{\p,A}^{*} \bw_{\p,B}^{\H}\bE_M } \ba_{2}(\psi) \right. \nonumber \\
	&\quad \left. + \ba_{1}^{\H}(\psi) \brc{\bx^{*} \by^{\T}} \otimes \brc{ \bw_{\p,B}^{*} \bw_{\p,A}^{\H}\bE_M } \ba_{2}(\psi) \right] G_0\brc{\phi,\theta} \nonumber\\
	&= 0.
\end{align}
Thus, the last term in~\eqref{eqn:totalPatternGolayUla} disappears and the radiated electric fields of the two subarrays add up incoherently at the receiver. Therefore, we have
\begin{align}
	G & \brc{\phi, \theta} \nonumber\\
	&= \norm{\be_{1}\brc{\phi, \theta}}^2 + \norm{\be_{2}\brc{\phi, \theta}}^2 \label{eqn:golayFieldsOrthogonalUla} \\
	&= G_0\brc{\phi,\theta} \!\left(\abs{ \brc{\bx \otimes \bw_{\p,A}}^{\T} \!\ba_{1}(\psi) }^2 \!\!\!+\!\abs{ \brc{\bx \otimes \bw_{\p,B}}^{\T} \!\ba_{1}(\psi) }^2 \right. \nonumber\\
	& \; \left.+\!\abs{ \brc{\by \otimes \bw_{\p,A}^* \bE_M}^{\T} \!\ba_{2}(\psi) }^2 \!\!\!+\!\abs{ \brc{\by \otimes \bw_{\p,B}^* \bE_M}^{\T} \!\ba_{2}(\psi) }^2\!\right)\!. \label{eqn:golayPatternsOrthogonalUla}
\end{align}
Next, we note that the array factor of the first part of the array, $\ba_1\brc{\psi}$, can be rewritten as
\begin{align}
	\ba_1\brc{\psi} = \ba_M\brc{M\psi}\otimes \ba_{\p}\brc{\psi},
\end{align}
where $\ba_M\brc{M\psi}$ is the array factor of a ULA of $N$ elements separated by $M d_y$ wavelengths apart. Using this, as well as the facts that 
$\ba_{2}\brc{\psi} = \e^{\j NM \psi}\; \ba_{1}\brc{\psi}$ and
\begin{align}
    \bE_{M}\ba_{\p}\brc{\psi} = \e^{\j\psi(M-1)} \ba_{\p}^*\brc{\psi},
\end{align} after regrouping the terms in~\eqref{eqn:golayPatternsOrthogonalUla}, we get
\begin{align}
\label{eqn:golayPowerPatternMultiplicationUla}
	G& \brc{\phi, \theta} = \brc{\abs{\bx^{\T}\ba_M\brc{M\psi}}^2 + \abs{\by^{\T}\ba_M\brc{M\psi}}^2} \nonumber \\
	&\;\;\times \brc{ \abs{\bw_{\p,A}^{\T}\ba_{\p}\brc{\psi}}^2 + \abs{\bw_{\p,B}^{\T}\ba_{\p}\brc{\psi}}^2} G_0\brc{\phi,\theta},
\end{align}

Next, $(\bx,\by)$ being a complementary pair, Prop.~\ref{prop:golayAsiEquivalence} yields
\begin{align}
\label{eqn:golayProtoarrayPreservationUla}
	G \brc{\phi, \theta} = 2N \;G_{\p} \brc{\phi, \theta},
\end{align}
Hence, the pattern of the entire array is a scaled version of the protoarray's pattern, meaning that the latter is preserved during the expansion.
\end{IEEEproof}

It is interesting to note that~\eqref{eqn:golayPowerPatternMultiplicationUla} represents the pattern multiplication property for a \emph{nested array}, consisting of subarrays that can be treated as virtual antenna elements. Weights $\brc{\bx,\by}$ could thus be referred to as \emph{expanders} for the protoarray weight pair $(\bw_{\p,A}, \bw_{\p,B})$, preserving the pattern produced by the latter. Moreover, if weights $(\bw_{\p,A}, \bw_{\p,B})$ also form a complementary pair, the entire pattern becomes
\begin{align}
\label{eqn:golayBroadBeam}
	G \brc{\phi, \theta} = 4NM \;G_0 \brc{\phi, \theta},
\end{align}
hence forming a beam as broad as the element pattern. Note also that the proposed construction constitutes a generalization of the ASI method presented in~\cite{petersson2020efficient}, the latter being a particular case of $\bx=\by=1$. Also, similarly to~\cite{petersson2020power}, this array enlargement can be conducted recursively, generalizing to $k$-fold expansion.

If a Golay pair cannot be constructed from pairs of shorter lengths, the former is referred to as a \emph{kernel} Golay pair~\cite{borwein2004complete}. Using the above constructions and a list of known quaternary Golay kernels~\cite[Sec.~4]{holzmann1994computer} it is possible to expand a protoarray of size $D$ to lengths $M = 2^{a+f+1} 3^b 5^c 11^d 13^e D$, where $f \leq c+e,\; b+c+d+e \leq a+2f+1,\; a,b,c,d,e,f \in \natural{}_0$, while preserving its pattern. And if the latter is a power-efficient broad beam, so will be the pattern of the entire array.

Although the list of known Golay sequences is rather extensive (see~\cite[Tab. 2]{fiedler2013small}), it is not complete. There are gaps in the lengths of known Golay sequences already for moderate, practically relevant, antenna array sizes, \eg, $M\in\{7,9,14,15,17,19,21\}$. Furthermore, general sufficient conditions for a pair of sequences to be a Golay pair remain unknown to date. Thus, it is of great interest to find polyphase Golay sequences for the missing lengths. 

\section{Search for New Complementary Sequences}
\label{sec:epsComplSequences}

Clearly, to fill the aforementioned gaps, one has to go beyond the quaternary alphabet and consider general polyphase sequences. Unfortunately, though, it is unclear how to search for polyphase sequences. Because the entire unit circle is considered as an alphabet, it is quite difficult to conduct an exhaustive search. The complexity of such a serch would grow exponentially with the sequence length, as well as with the quantization granularity of interest. Notably, unsuccessful searches for octernary sequences of lengths 7 and 9 were reported in~\cite{ghaderpour2013asymptotic}. Also, certain results on hexternary Golay sequences are available in~\cite{fiedler2013small}.

\subsection{$\epsilon$-Complementary Sequences}
In contrast to an exhaustive search, here we take a different route in this paper, addressing the problem by means of stochastic optimization. For that purpose, similarly to~\cite{qiao2015broadbeam}, we relax the requirements on spatially flat array factor in~\eqref{eqn:defBroadBeam} by introducing certain angular fluctuation. From the complementarity viewpoint, this is equivalent to allowing for a certain level of imperfection in the sum AACF for the sequence pair of interest. Let us define the following concept.
\begin{definition}[$\epsilon$-complementary pair]
	A pair of unimodular sequences $(\bu,\bv)$ is called an $\epsilon$-complementary pair if the following holds
	\begin{equation}
		\label{eqn:defEpsComplSequences}
		\abs{R_{\bu}(\tau) + R_{\bv}(\tau)} \leq \epsilon, \qquad \forall \tau \neq 0,
	\end{equation}
	where $\epsilon$ is some small tolerance threshold on the level of side lobes of the sum AACF. 
\end{definition} 

When $\epsilon$-complementary sequences are used for beamforming, property~\eqref{eqn:defEpsComplSequences} translates into a certain amount of ripple in the sum power spectra. However, the level of the ripple is guaranteed to be limited. This relaxation enables the usage of various heuristics to obtain beams practically indistinguishable from broad beams.

\subsection{Modified Great Deluge Algorithm}
To find the said $\epsilon$-complementary sequence pairs we next develop a corresponding heuristic algorithm. Our starting point is the \emph{Great Deluge algorithm} (GDA), invented by D\"{u}ck in~\cite{dueck1993new}, and successfully applied for numerical searches for polyphase Barker sequences~\cite{friese1994polyphase}. The GDA has been shown to outperform other popular stochastic-optimization algorithms, such as simulated annealing~\cite{kirkpatrick1983optimization} and threshold accepting~\cite{dueck1990threshold} for certain classes of optimization problems~\cite{dueck1993new}.

Herein, we propose the \emph{modified GDA} (MGDA) algorithm, adapted to operation with a pair of polyphase sequences. The aim of the proposed MGDA is to optimize the phases of two phase sequences $\bphi_{\bu} \triangleq \sqbrc{ \varphi_{\bu,1},\ldots,\varphi_{\bu,M}}^{\T}$ and $\bphi_{\bv} \triangleq \sqbrc{\varphi_{\bv,1},\ldots,\varphi_{\bv,M}}^{\T}$, minimizing the level of side lobes of their sum AACF. The phases are related to the sequences through 
\begin{align}
\label{eqn:mapPhasesWeights}
	u_{m} = \e^{\j\varphi_{\bu,m}}, \quad v_{m} = \e^{\j\varphi_{\bv,m}}, \quad \forall m\in\{1,\ldots,M\}.
\end{align}
For later convenience, we define the combined vector of phases, stacking the phase vectors on top of each other, $\bphi \triangleq [\bphi_{\bu}^{\T}, \bphi_{\bv}^{\T}]^{\T} \in [0, 2\pi)^{2M}$. 

The objective of the MGDA is the maximization of a certain utility function associated with the given phase vector, which can be formulated as follows
\begin{equation}
    \begin{aligned}
        &\underset{\bphi}{\maximize} 
        & & U\brc{\bphi}\\
        &\st 
        & &\bphi \in [0, 2\pi)^{2M},
        \end{aligned}
\end{equation} 
where $U\brc{\bphi}$ is a utility function accounting for the side lobe level. To satisfy~\eqref{eqn:defEpsComplSequences}, we select
\begin{align}
\label{eqn:defUtility}
    U\brc{\bphi} = -\max_{\substack{\tau\in \{-M+1,\ldots,M-1\}, \\ \tau \neq 0}}\figbrc{\abs{ R_{\bu}\brc{\tau} + R_{\bv}\brc{\tau} }}.
\end{align}

Ideally, the side lobes of the sum AACF should vanish altogether (\ie, $U\brc{\bphi} = 0$), as a result of the optimization. However, keeping their level within some tolerance interval $\epsilon$ around zero might suffice for practical beamforming purposes.

\begin{algorithm}
\caption{Proposed MGDA for searching for polyphase $\epsilon$-complementary sequence pairs.}
\label{alg:mgda}
	\begin{algorithmic}[1]
		\REQUIRE 
			{Rain intensity $V>0$, phase scaling factor $\alpha \in (0,1]$, tolerance threshold $\epsilon>0$, maximum number of unsuccessful steps $d_{\max}$.}
		\STATE Initiate phase vector $\bphi \sim \calU[0, 2\pi)^{2M}$.
		\STATE Initiate phase increments vector $\Delta\bphi \sim \calU[0, 2\pi)^{2M}$.
		\STATE Set the water level: $\lambda \leftarrow  U\brc{\bphi}$.
		\STATE Initialize the number of unsuccessful steps: $d \leftarrow 0$.
		    		
		\texttt{// Zooming-in local exploration}
        \WHILE{$U\brc{\bphi} > \epsilon$}
    		\FOR{$i = 1 \to 2M$}
    		    \WHILE{$U\brc{\bphi} < \lambda$}
    		   		\STATE Increment phase: $\varphi_i \leftarrow \varphi_i + \Delta\varphi_i$.
            		\IF{$U(\bphi)\geq\lambda$}
            		    \STATE Accept the new phase $\varphi_i$
            		\ELSE
                		\STATE Step backwards: $\varphi_i \leftarrow \varphi_i - \Delta\varphi_i$.
                		\IF{$U(\bphi)\geq\lambda$}
                		    \STATE Accept the new phase $\varphi_i$
                		\ELSE
                		    \STATE Scale down the step size: $\Delta\varphi_i \leftarrow \alpha \Delta\varphi_i$.
                		\ENDIF
            		\ENDIF
        		\ENDWHILE
    		\ENDFOR
    		
		    \texttt{// Flooding}
		    \STATE Increase the water level: $\lambda \leftarrow \lambda + V$.
    		
		    \texttt{// Zooming-out global exploration}
    		\IF{$d \geq d_{\max}$}
        		\STATE Scale up the random step size: $\Delta\bphi \sim \calU[0, 2\pi)^{2M}$.
        		\STATE Jump to a new neighborhood: $\bphi \leftarrow \bphi + \Delta\bphi$.
        		\STATE Reset the water level: $\lambda \leftarrow  U\brc{\bphi}$.
        	    \STATE Reset the number of unsuccessful steps: $d \leftarrow 0$.
    		\ENDIF
		\ENDWHILE
		\RETURN Optimized phase vector $\bphi$.
	\end{algorithmic}
\end{algorithm}    

\begin{figure}
	\centering
	\begin{subfigure}{0.495\textwidth}
		\centering
		\includegraphics[width=8cm]{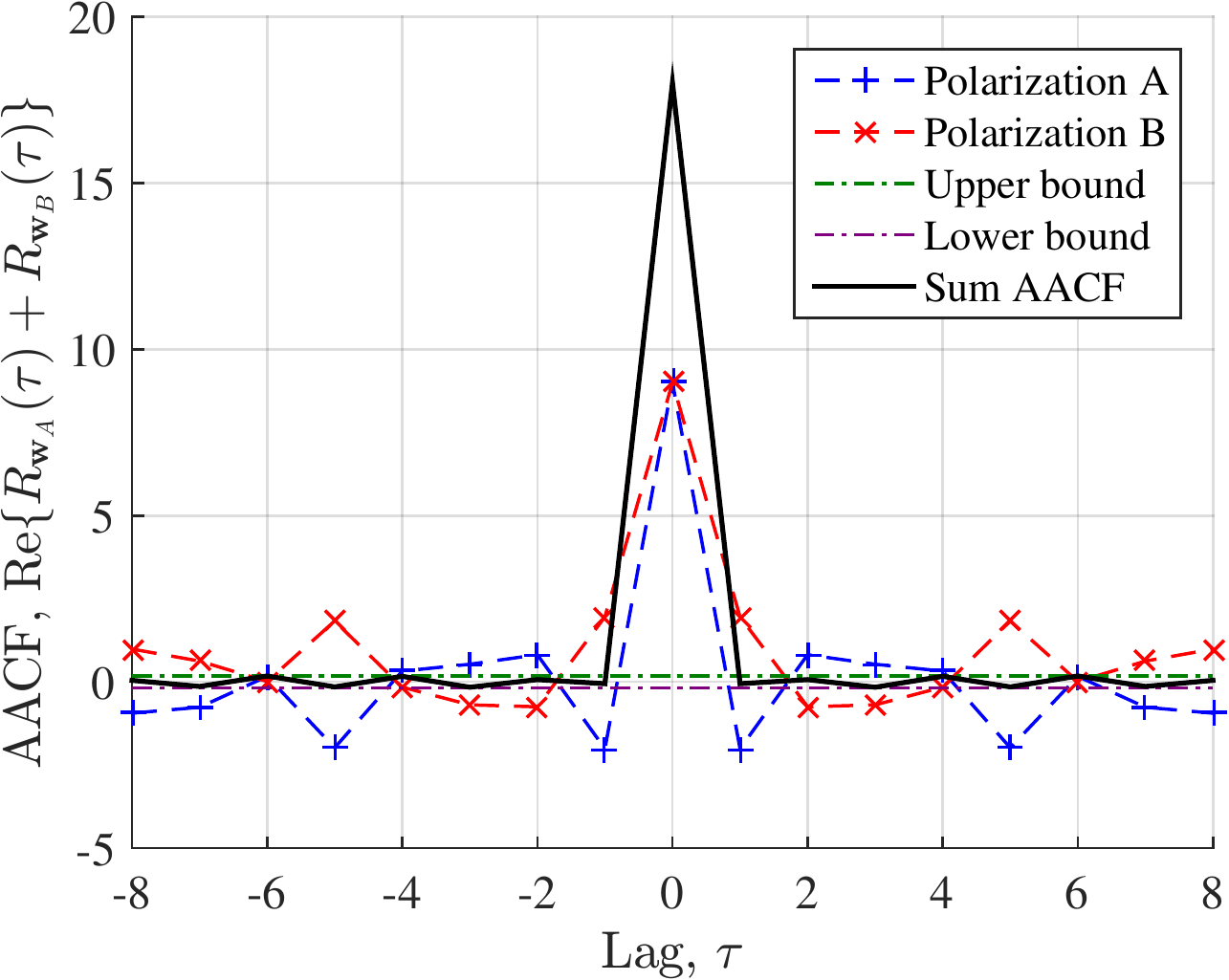}
		\caption{Aperiodic autocorrelation function.}
		\label{fig:plotAacfExample}
	\end{subfigure}
	\begin{subfigure}{0.495\textwidth}
		\centering
		\includegraphics[width=8cm, trim = 0cm 0cm 0cm -0.5cm, clip=True]{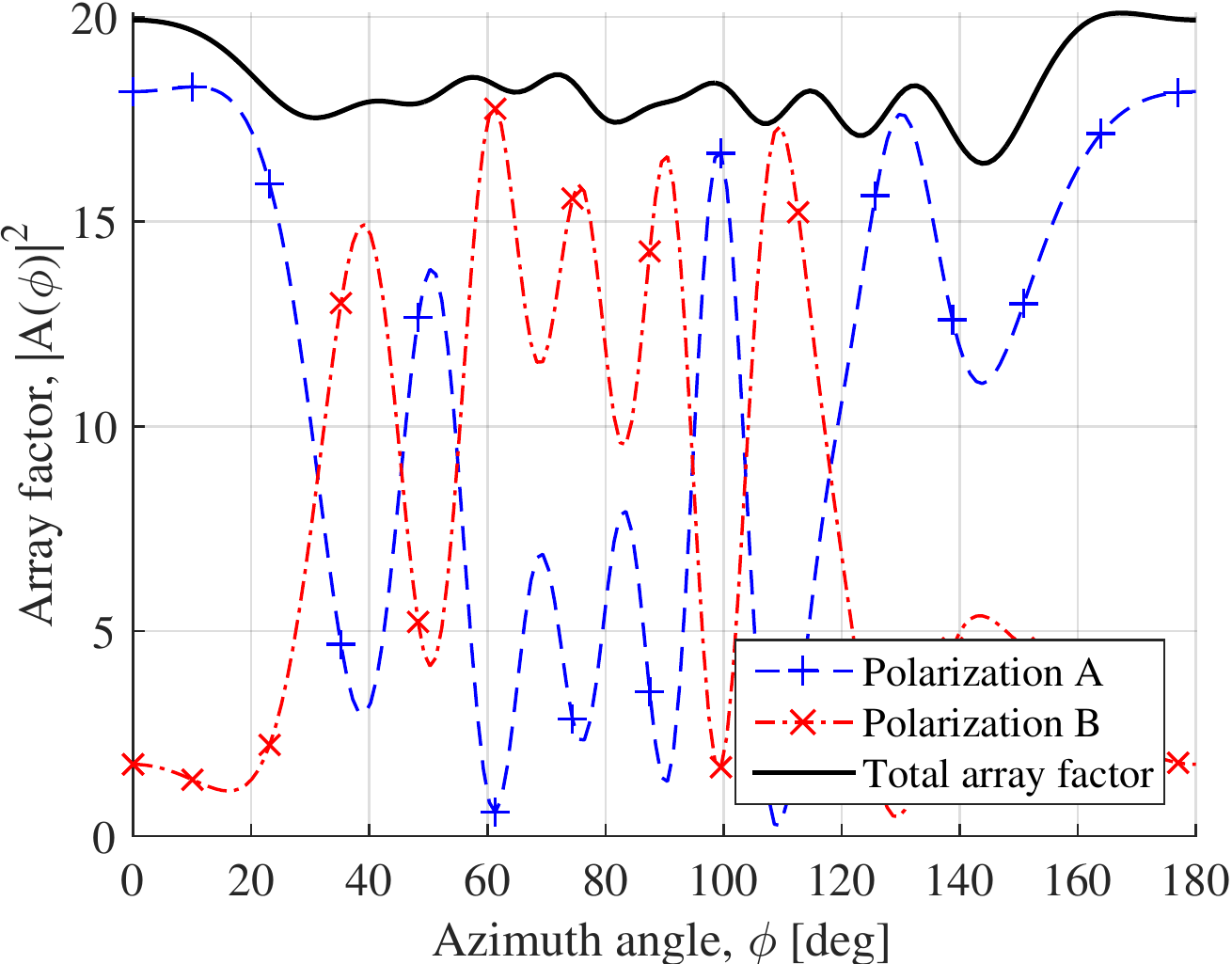}
		\caption{Power-domain array factor.}
		\label{fig:plotAfExample}
	\end{subfigure}
	\caption{Illustration of the sum-AACF and array factor of a pair of $\epsilon$-complementary vectors $\bw_A$ and $\bw_B$ for a horizontal ULA with $M = 9$ antennas.}
	\label{fig:plotAacfAfExample}
	\vspace{-0.5cm}
\end{figure}

The implementation details of the proposed MGDA are summarized in Alg.~\ref{alg:mgda}. The corresponding stochastic optimization process describes the movement of a current state point over a landscape formed by the utility function. In contrast to gradient ascent, which may move only uphills, Alg.~\ref{alg:mgda} is allowed to move the current state both uphills and downhills, as long as it is above a certain \emph{water level}. This prevents the MGDA from being quickly stuck at a local maximum. During the optimization, the water level keeps increasing, forcing the algorithm to escape into areas of higher utility, and eventually end up near a (hopefully global) maximum.

The implementation of the above principle is done by increasing each phase $\varphi_i$ by a corresponding increment value $\Delta\varphi_i$. Then, the resulting value of the utility function is examined. If the utility value is above the water level, the new state is accepted and the next phase is altered; otherwise, the phase step is taken in the opposite direction and the corresponding utility value is compared to the water level. If no increase is accepted in any of the directions, the step size is decreased by a factor $\alpha<1$ and the procedure is repeated until an eligible step is accepted. 

In addition to the above, large-scale exploration is added to the algorithm. Namely, we track whether the number of unsuccessful alternations is greater than a certain number $d_{\max}$. If so, the system takes a large step to a new random configuration, and the entire process restarts. The algorithm is terminated when the utility function reaches the neighborhood of the known optimum, $U\brc{\bphi} = 0$, as determined by the adopted tolerance level $\epsilon$. 

For the sake of illustration, Fig.~\ref{fig:plotAacfExample} plots the imperfect sum-AACF and the power-domain array factor of an $\epsilon$-complementary pair of beamforming vectors of size $M=9$, obtained by means of Alg.~\ref{alg:mgda}. It can be seen from the figure that the side lobes of the sum AACF lie within the upper ($+\epsilon$) and lower ($-\epsilon$) bounds defined by tolerance which, for this example, was chosen as 1\% of the main lobe level of the sum-AACF, \ie, $\epsilon=0.02 M=0.18$. Fig.~\ref{fig:plotAfExample}, in turn, plots the corresponding array factor. Note how the per-polarization beams exhibit different behaviors, while adding up into a smooth and broad shape. Observe also certain ripple in the array factor due to non-zero side lobes of the sum AACF.\footnote{Note that a rather large tolerance threshold was chosen to highlight the imperfection of the obtained solution.}

\section{Two-Dimensional Arrays}
\label{sec:broadBeamsUpa}

In Section~\ref{sec:broadBeamsArbitraryUla}, we carried out the expansion of a one-dimensional dual-polarized ULA. Here, let us consider a further generalization of this process to a two-dimensional dual-polarized URA. 

\begin{figure*}[!t]
\normalsize
\setcounter{equation}{46}
\begin{align}
\label{eqn:defAacf2d}
    R_{\bU}(\tau_n,\tau_m) = \left\{ 
        \begin{matrix} 
        &\sum\limits_{n=1}^{N-\tau_n} \sum\limits_{m=1}^{M-\tau_m} u_{n,m} u_{n+\tau_n,m+\tau_m}^*, 
        & & 0\leq \tau_n \leq N-1,\enspace 0\leq \tau_m \leq M-1,\\
        &\sum\limits_{n=1}^{N+\tau_n} \sum\limits_{m=1}^{M-\tau_m} u_{n-\tau_n,m} u_{n,m+\tau_m}^*, 
        & & -N+1\leq \tau_n < 0,\enspace 0\leq \tau_m \leq M-1,\\
        &\sum\limits_{n=1}^{N-\tau_n} \sum\limits_{m=1}^{M+\tau_m} u_{n,m-\tau_m} u_{n+\tau_n,m}^*, 
        & & 0\leq \tau_n \leq N-1,\enspace -M+1\leq \tau < 0,\\
        &\sum\limits_{n=1}^{N+\tau_n} \sum\limits_{m=1}^{M+\tau_m} u_{n-\tau_n,m-\tau_m} u_{n,m}^*, 
        & & -N+1\leq \tau_n < 0,\enspace -M+1\leq \tau < 0,\\
        & 0,
        & & \tau_n \notin (-N, N), \enspace \tau_m \notin (-M, M)
        \end{matrix} \right. 
\end{align}
\setcounter{equation}{47}
\hrulefill
\end{figure*}

\begin{definition}[Two-dimensional AACF]
    Let $\bU \in \complex{N\times M}$ be a two-dimensional array/matrix of complex unit-norm values. Its two-dimensional AACF is defined as in~\eqref{eqn:defAacf2d} on the top of the next page, where $u_{n,m}$ denotes the $(n,m)$th entry of a matrix~$\bU$.
\end{definition}

Based on the above, a two-dimensional complementary array pair is defined as follows~\cite{ohyama1978advanced}.
\begin{definition}[Golay array pair]
    A pair of unimodular arrays $(\bU,\bV)$ is called a Golay complementary array pair if the following holds
    \begin{align}
    \label{eqn:defGolayArray}
        R_{\bU}(\tau_n,\tau_m) + R_{\bV}(\tau_n,\tau_m) &= 2NM \delta (\tau_n,\tau_m).
    \end{align}
    Moreover, an array $\bU\in\complex{N\times M}$ that forms a Golay pair with another array $\bV\in\complex{N\times M}$ is referred to as a Golay array. 
\end{definition}

Given a URA size, finding a complementary pair of arrays provides excitation matrices that produce a broad beam. Below, we present several recursive constructions for such arrays.

\subsection{ULA to URA Expansion}

First, we provide a construction of a two-dimensional array from a ULA. Having constructed a linear protoarray with a broad beam, we can expand it to form a URA that preserves its pattern.
\begin{proposition}
\label{prop:weights2d}
	Assume that a one-dimensional protoarray of size $M$, excited with beamforming weights $(\bw_{\p,A}, \bw_{\p,B})$, produces a desired radiation pattern. Furthermore, let  $(\bx, \by)$ be a polyphase Golay sequence pair of length $N$. Then the expanded two-dimensional array with excitation weight matrices given either by
	\begin{subequations}
	\label{eqn:weights2dVert}
		\begin{align}
    	\bW_A &= \sqbrc{
    		\begin{matrix}
    		\bx\bw_{\p,A}^{\T}\\
    		-\by\bw_{\p,B}^{\H}\bE_M
    		\end{matrix}
    	} \in \complex{2N\times M},\\
    	\bW_B &= \sqbrc{
    		\begin{matrix}
    		\bx\bw_{\p,B}^{\T}\\
    		\phantom{-}\by\bw_{\p,A}^{\H} \bE_M
    		\end{matrix}
    	} \in \complex{2N\times M},
    	\end{align}
  \end{subequations}
    
    or, alternatively, by
    \begin{subequations}
    \label{eqn:weights2dHoriz}
		\begin{align}
    	\bW_A &= \sqbrc{
    		\begin{matrix}
    		\bx\bw_{\p,A}^{\T}, \;
    		-\by\bw_{\p,B}^{\H}\bE_M
    		\end{matrix}
    	}\in \complex{N\times 2M},\\
    	\bW_B &= \sqbrc{
    		\begin{matrix}
    		\bx\bw_{\p,B}^{\T}, \;
    		\phantom{-}\by\bw_{\p,A}^{\H} \bE_M
    		\end{matrix}
    	}\in \complex{N\times 2M}
    	\end{align}
    \end{subequations}
	preserves the radiation pattern of the protoarray.
\end{proposition}

\begin{IEEEproof}
    The proof is similar to that of Prop.~\ref{prp:golayConstructionUla}. The details are skipped for brevity. In a nutshell, we show that 
    \begin{align}
    \label{eqn:golayPowerPatternMultiplicationUpa}
    	G& \brc{\phi, \theta} = \brc{\abs{\bx^{\T}\ba_z\brc{\psi_z}}^2 + \abs{\by^{\T}\ba_z\brc{\psi_z}}^2} \nonumber \\
    	&\times \brc{ \abs{\bw_{\p,A}^{\T}\ba_{\p}\brc{\psi_y}}^2 + \abs{\bw_{\p,B}^{\T}\ba_{\p}\brc{\psi_y}}^2} G_0\brc{\phi,\theta},
    \end{align}
    The pair $(\bx,\by)$ being a complementary pair, we obtain
    \begin{align}
    \label{eqn:golayProtoarrayPreservationUpa}
    	G \brc{\phi, \theta} = 2N \;G_{\p} \brc{\phi, \theta},
    \end{align}
    and hence the pattern of the protoarray is preserved.
\end{IEEEproof}

This result is related to the construction reported in~\cite{petersson2020power}. Note that the obtained arrays must have \emph{even} number of elements in at least one of the two dimensions.

Based on the results so far, broad beams can be designed for various array configurations. For the configurations for which no Golay sequence pair is known, the latter can be approximated by $\epsilon$-complementary pairs. These can be further expanded to two-dimensional arrays of various sizes.

\subsection{Expansion of URAs}

Having constructed a two-dimensional array with desired radiation properties, we can expand it to larger sizes via the following construction.

\begin{proposition}
	\label{prop:weightsExpansionUpa}	
	Assume that a two-dimensional protoarray of size $N\times M$, excited with beamforming weights $(\bW_{\p,A}, \bW_{\p,B})$, exhibits a desired radiation pattern. Furthermore, let  $(\bX, \bY)$ be a polyphase Golay array pair of size $L\times K$. An expanded array formed either by
	\begin{subequations}
	\label{eqn:expansionVertical}
		\begin{align}
		\bW_A &= \sqbrc{
			\begin{matrix}
			\bX\otimes\bW_{\p,A}\\
			-\bY\otimes\brc{\bE_N\bW_{\p,B}^{*}\bE_M}
			\end{matrix}
		}\in\complex{2LN\times KM},\\
		\bW_B &= \sqbrc{
			\begin{matrix}
			\bX\otimes\bW_{\p,B}\\
			\phantom{-}\bY\otimes\brc{\bE_N\bW_{\p,A}^*\bE_M}
			\end{matrix}
		}\in\complex{2LN\times KM},
		\end{align}
	\end{subequations}
	or, alternatively, by
	\begin{subequations}
		\begin{align}
		\bW_A \!&=\! \sqbrc{
			\begin{matrix}
			\bX\otimes\bW_{\p,A},\;
			-\bY\otimes\brc{\bE_N\bW_{\p,B}^{*}\bE_M}
			\end{matrix}
		}\in\complex{LN\times 2KM},\\
		\bW_B \!&=\! \sqbrc{
			\begin{matrix}
			\bX\otimes\bW_{\p,B},\;
			\phantom{-}\bY\otimes\brc{\bE_N\bW_{\p,A}^*\bE_M}
			\end{matrix}
		}\in\complex{LN\times 2KM},
		\end{align}
	\end{subequations}
	preserves the radiation pattern of the protoarray.
\end{proposition}

\begin{IEEEproof}
    The proposition is a generalization of Prop.~\ref{prop:weights2d}, and its proof is analogous. The details are skipped for brevity.
\end{IEEEproof}

Again, we note that at least one of the dimensions of the obtained arrays must be even. To deal with arrays that have both dimensions odd, we develop the following concept.

\subsection{$\epsilon$-Complementary Arrays}

Extending the notion introduced in Sec.~\ref{sec:epsComplSequences}, with aim of enlarging the practically-feasible set of available beamforming matrices, we define $\epsilon$-complementary arrays as follows.
\begin{definition}[$\epsilon$-complementary array pair]
    A pair of unimodular arrays $(\bU,\bV)$ is called an $\epsilon$-complementary array pair if the following holds
    \begin{align}
    \label{eqn:defEpsComplArray}
        |R_{\bU}(\tau_n,\tau_m) + R_{\bV}(\tau_n,\tau_m)| &\leq \epsilon, & \forall \tau_n\neq 0, \tau_m\neq 0,
    \end{align}
    where $\epsilon$ is some small tolerance threshold on the level of side lobes of the sum AACF.  
\end{definition}

The above $\epsilon$-complementary arrays can be obtained by direct application of the proposed MGDA. However, a slight modification is needed: the two arrays in a pair shall be vectorized prior to being stacked into a phase vector, \ie,
\begin{align}
    \bphi = \sqbrc{\vec\brc{\bphi_{\bU}}^{\T}, \vec\brc{\bphi_{\bV}}^{\T}}^{\T}.
\end{align}

Unfortunately, in the two-dimensional setting with both dimension odd, the MGDA does not seem to converge to an $\epsilon$-complementary array pair for $\epsilon \lesssim 3\%$ of the main sum-AACF lobe. It is not clear whether there is a fundamental limit on the level of side lobes of $\epsilon$-complementary array pairs or whether our experimental runs of the MGDA were unlucky in finding the global optimum. Nevertheless, as indicated by several runs, for the configurations for which there is a known Golay array pair, the algorithm did converge to a solution with arbitrarily small $\epsilon$. This hints at decent performance of the MGDA in the two-dimensional setting, also at potential nonexistence of Golay arrays with both sizes odd. Notwithstanding,  gaps remain in feasible array sizes and no methods are known to date that are able to tackle this problem for those configurations. This aspect, thus, remains a topic for future work.

\section{Numerical Examples}
\label{sec:numericalExamples}

For the sake of illustration, this section provides several numerical examples, comparing the proposed approach to the existing broad-beamforming techniques described in Sec.~\ref{sec:broadBeamDesign}.\footnote{The codes producing the figures in this paper are available at:\\ \texttt{https://github.com/girnyk/GolayBeamforming}.}

\subsection{Broad-Beam Design for ULA}
\label{sec:simBroadBeamUla}
Consider a BS equipped with a ULA containing $M=7$ antenna elements\footnote{Note that such an antenna configuration is encountered in real-world systems. For example, the widely-used BS antenna Kathrein~800~10204 is equipped with 7 radiating elements~\cite[p. 24]{kathrein2014694}.} with maximum gain of 8 dBi and element separation of $d_{y}=0.5$ wavelengths. As we have seen before, no Golay sequences are known for this length, and hence, conventional methods cannot create a broad beam with full power utilization for such configuration. Instead, they typically result either in taper loss and/or radiation pattern ripple. To obtain the excitation weights for this array layout we apply the MGDA presented in Alg.~\ref{alg:mgda} with side lobe tolerance $\epsilon$ set to $1\%$ of the level of the main sum-AACF lobe, \ie, $\epsilon=0.02M=0.14$. This results in the following pair of weights
\begin{subequations}
\label{eqn:weightsEpsCompl7}
	\begin{align}
	\bw_A &= \exp(\j \; [0, 0.97, 1.83, 4.98, 0.16, 3.34, 1.20]^{\T}),\\
	\bw_B &= \exp( \j \;[0, 1.75, 0.86, 2.21, 1.12, 5.75, 4.41]^{\T}).
	\end{align}
\end{subequations}

\begin{figure}
	\centering
	\includegraphics[width=8cm]{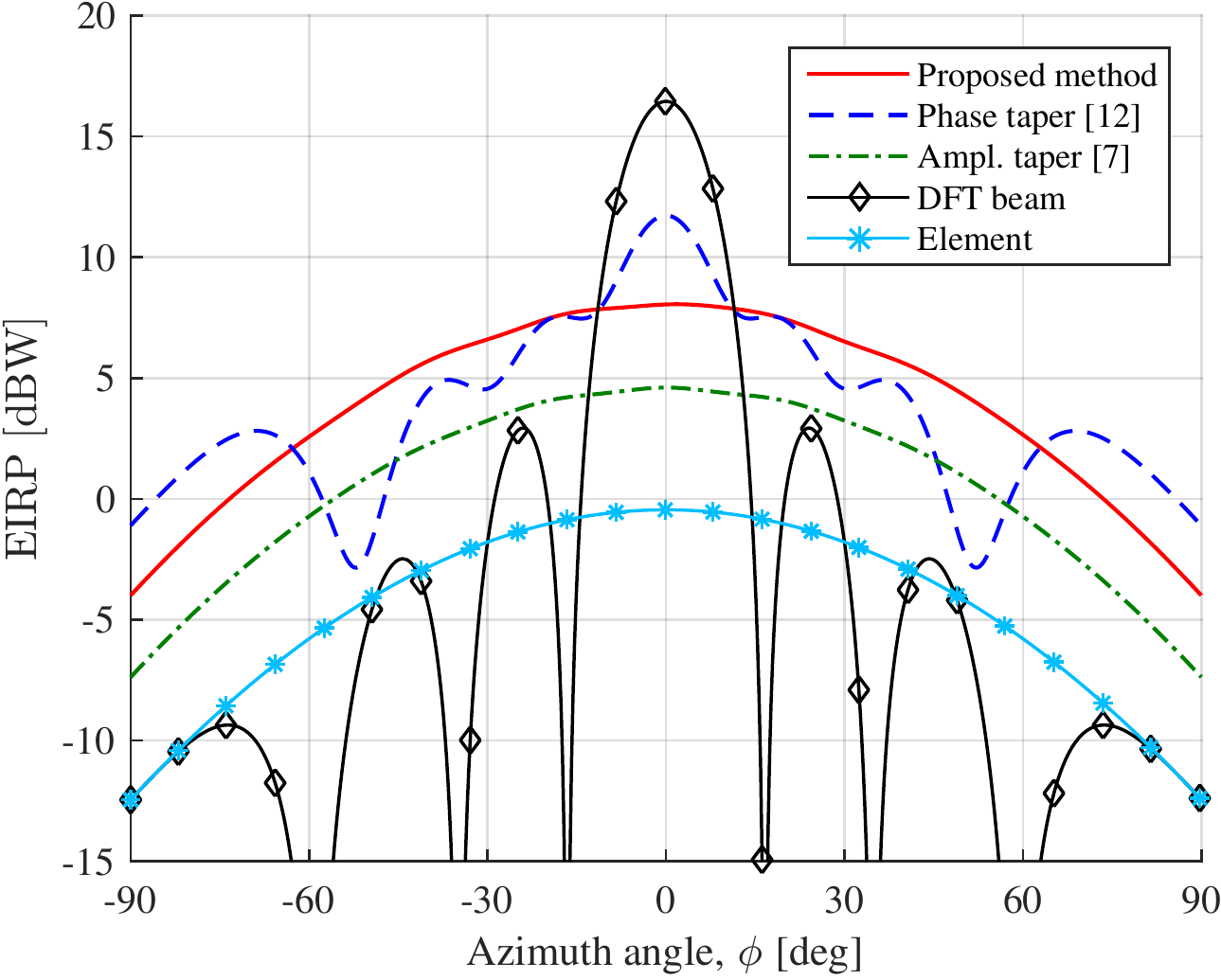}
	\caption{Effective isotropic radiated power patterns for a horizontal ULA with $M=7$ antennas.}
	\label{fig:plotEirpUla}
  	\vspace{-0.3cm}
\end{figure}

For illustration sake, Fig.~\ref{fig:plotEirpUla} compares the power radiation patterns of the proposed beamforming method---with excitation weights~\eqref{eqn:weightsEpsCompl7}---and the other considered methods in terms of equivalent isotropically radiated power (EIRP). That is, the beamforming weights are normalized, such that the total output power is 1~W, minus eventual loss from amplitude tapering.\footnote{Practically, this is done by normalizing the weights with $1/\brc{\sqrt{2M}\max\figbrc{\abs{[\bw_A^{\T}, \bw_B^{\T}]^{\T}}}}$.} The amplitude tapering weights in~\eqref{eqn:weightsQiao} are optimized with tolerance $\zeta=0.01$. The weights for the phase taper are picked from~\eqref{eqn:weightsIntel} with $p=3$ and $c=24$. The figure also plots the radiation pattern of an individual element, given by the commonly adopted 3GPP subelement model~\cite{3gpp2017study}
\begin{equation}
\label{eqn:subelementPattern3gpp}
    G_0\brc{\phi} = 8 - \min\figbrc{12\brc{ \frac{\phi-\phi_0}{\Delta \phi}}^2, 30} \quad [\mathrm{dB}],
\end{equation}
where the pointing direction is set to be $\phi_0 = 0^{\circ}$, whereas the HPBW is chosen as $\Delta \phi = 90^{\circ}$. 


From the figure, we see that the proposed beam is as broad as the pattern of a subelement, while exhibiting higher radiation level and showing the best balance of power utilization and ``broadness'' of the beam. The proposed weights are furthermore unimodular, and hence possess excellent PA utilization. Thus, the proposed approach efficiently solves the task impossible to handle by the conventional methods for the given array configuration.

\subsection{Performance Results}
To evaluate the performance of the proposed method, we consider a single $120^{\circ}$-wide sector of radius 300m for which we assume line-of-sight conditions and single-layer transmission from a BS to a UE. Consider $K=10\,000$ UE locations uniformly distributed within the sector, \ie, $\phi_k \sim \calU[-60^{\circ}, 60^{\circ}]$ and $d_k \sim \sqrt{\calU[25\textrm{m}, 300\textrm{m}]}, \forall k \in\figbrc{1,\ldots,K}$, where $\phi_k$ is the azimuth direction and $d_k$ is the distance from the BS of UE location $k$. Ignoring shadow fading, the spectral efficiency for the UE at location $k$ is given by
\begin{equation}
\label{eqn:spectralEfficiency}
    C_k = \log_2\sqbrc{1 + \rho\gamma\,G(\phi_k)\,d_k^{-\alpha}},
\end{equation}
where $G(\phi_k)$ is the antenna gain in the direction of that UE location, normalized such that the total output power is 1 (minus taper loss). Meanwhile, $\rho = 1 / \sigma^2$ is the SNR, $\sigma^2$ being the receiver noise variance. 
Moreover, $\alpha=2.2$ is the pathloss exponent, and $\gamma=57$ dB is an offset that accounts for all other system gains and losses in the link budget, such as, \eg, output power, element gains, noise figure and distance-independent part of pathloss. We assume that the UEs are equipped with omnidirectional antennas, and hence the UE rotation does not impact the performance. The \emph{average spectral efficiency} is then computed as the arithmetic mean of~\eqref{eqn:spectralEfficiency} over the UE locations.

\begin{figure}
	\centering
	\includegraphics[width=7.8cm]{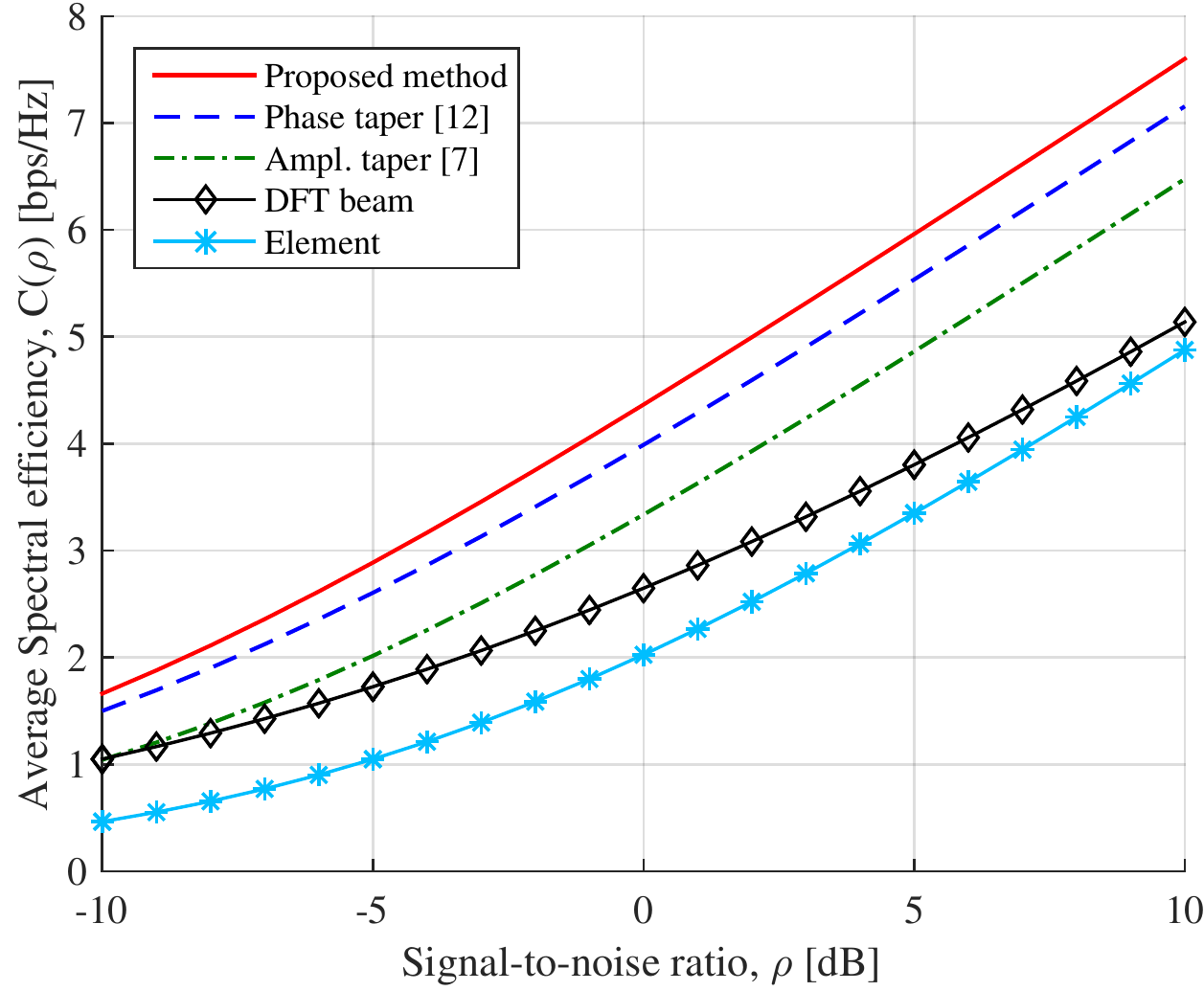}
	\caption{Average spectral efficiency achieved for a horizontal ULA with $M=7$ antennas.}
	\label{fig:spectralEfficiency}
	\vspace{-0.5cm}
\end{figure}

Fig.~\ref{fig:spectralEfficiency} plots the average spectral efficiency as a function of SNR for the methods under consideration. It can be seen that the proposed method outperforms the rest of the methods in terms of spectral efficiency. This is not surprising, given its previously shown superior performance in terms of power utilization and HPBW of the obtained radiation pattern. 

\begin{figure}[t]
	\centering
	\includegraphics[height=5.5cm]{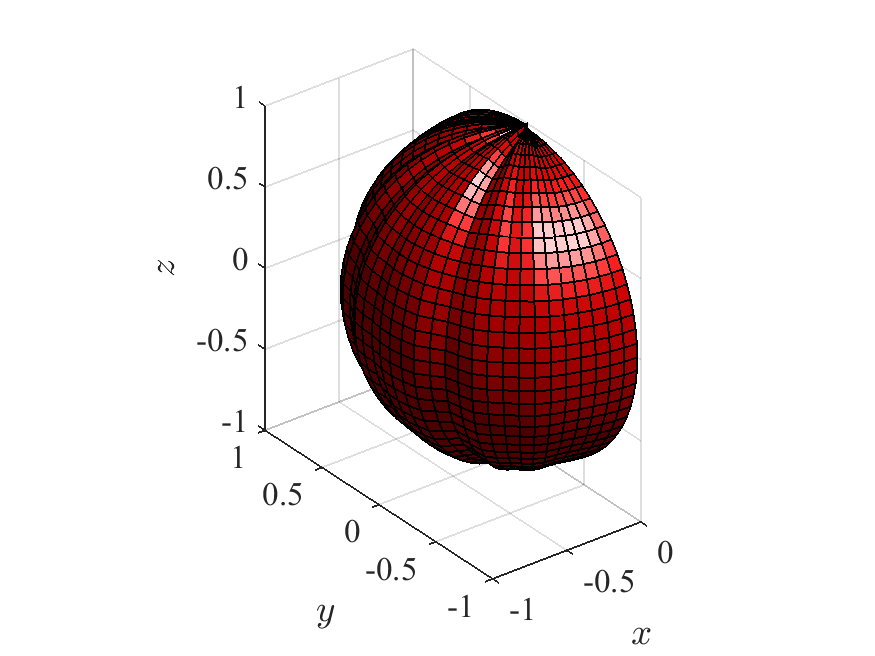}
	\caption{Normalized power-domain array factor of a URA of size $32\times 14$ excited with $\epsilon$-complementary weight matrices.}
	\label{fig:plotAf2d}
	\vspace{-0.3cm}
\end{figure}

\subsection{Broad-Beam Design for URA}
To illustrate the performance of ASI beamforming in a two-dimensional setting, we consider a large URA with $M=14$ vertical columns of $N=32$ cross-pole antennas each. The nearest-neighbor spacings are set to $d_{y}=d_{z}=0.5$ wavelengths in both dimensions. We place the array on a vertical $yz$-plane so that the array boresight direction is aligned with the $x$-axis. The weight matrices for both polarizations are obtained by expanding the pair of weights obtained in~\eqref{eqn:weightsEpsCompl7} by means of Alg.~\ref{alg:mgda}. These are used as horizontal weights and are expanded horizontally using~\eqref{eqn:weights2dHoriz} and then 3 times vertically using~\eqref{eqn:weights2dVert} with expander weights being $\bx=[1,1]^{\T}$ and $\by=[-1,1]^{\T}$. Fig.~\ref{fig:plotAf2d} shows the power-domain array factor of the beamformer. It is clearly seen from the figure that the array factor is close to being spatially flat, \ie, forming a semi-sphere. At the same time, as the weights are unimodular, full power utilization is achieved for all the antenna elements.

\section{Conclusions}
\label{sec:conclusions}

In this paper, we have proposed an efficient method for designing power-efficient cell-specific broad beams (\viz, beams designed with constant-modulus weights and characterized by spatially flat power-domain array factor) for public-channel transmission. The method is based on phase-only weight optimization and is, therefore, suitable for novel hybrid beamforming architectures. The method exploits the connection between dual-polarized beamforming and polyphase Golay complementary sequence. Based on this connection, we have proposed several ways for constructing larger arrays from smaller ones, while preserving the broad-beam property and constant-modulus weights. Unfortunately, there are gaps in the lengths of complex Golay sequences known to date, and hence in feasible array sizes. To fill these gaps, we have introduced the concept of $\epsilon$-complementarity that relaxes the strict requirement of Golay sequence pairs on zero side lobes of their sum AACF. Based on this relaxation, we have been able to develop the MGDA algorithm that finds $\epsilon$-complementary sequence pairs of a given length by means of stochastic optimization. This enables power-efficient broad beamforming for array configurations insupportable before. In addition, we have discussed the extension of the proposed method to two-dimensional rectangular arrays. Our approach provides means for broad beamforming for any two-dimensional array configuration except for those with odd number of antenna elements in both dimensions. Numerical illustrations show that the proposed approach outperforms state-of-the-art broad-beamforming methods.

\appendices

\section{Realistic Antenna Elements}
\label{sec:realisticElements}
As mentioned in Sec.~\ref{sec:ula}, cross-pole antennas must not be confused with two mechanically rotated crossed dipoles. Instead, real-world antenna elements are specifically designed to have a certain radiation pattern in an angular range of interest. From the practical viewpoint, one considers a \emph{service area} limited to a certain operational range of angles $\brc{\phi,\theta}$ that are served by the BS. For instance, for a typical three-sector deployment, $\phi \in [-60^{\circ}, 60^{\circ}]$ and $\theta \in [75^{\circ},115^{\circ}]$. Hence, both ports of an antenna element should exhibit similar radiation patterns, while maintaining orthogonal polarizations, within this service area (rather than for all possible angular directions).

\begin{figure}
	\centering
	\begin{subfigure}{0.5\textwidth}
		\centering
		\includegraphics[width=8cm, trim = 0.8cm 1cm 3cm 1cm, clip=True]{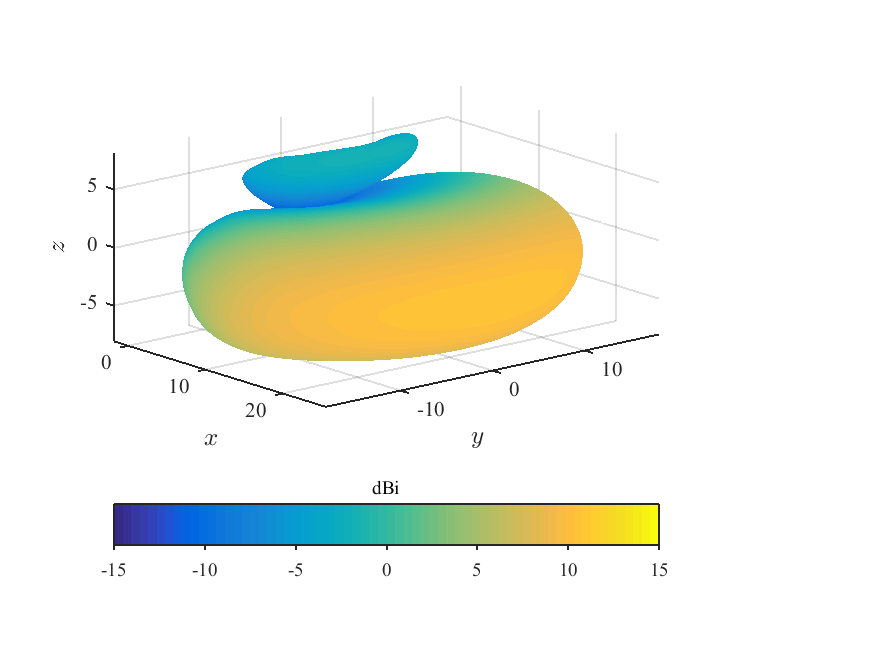}
    	\caption{Average gain over subarrays and polarizations.}
    	\label{fig:patternMeasSubarray}
	\end{subfigure}
	\begin{subfigure}{0.5\textwidth}
		\centering
		\includegraphics[width=8cm, trim = 0cm 0cm 0cm -0.5cm, clip=True]{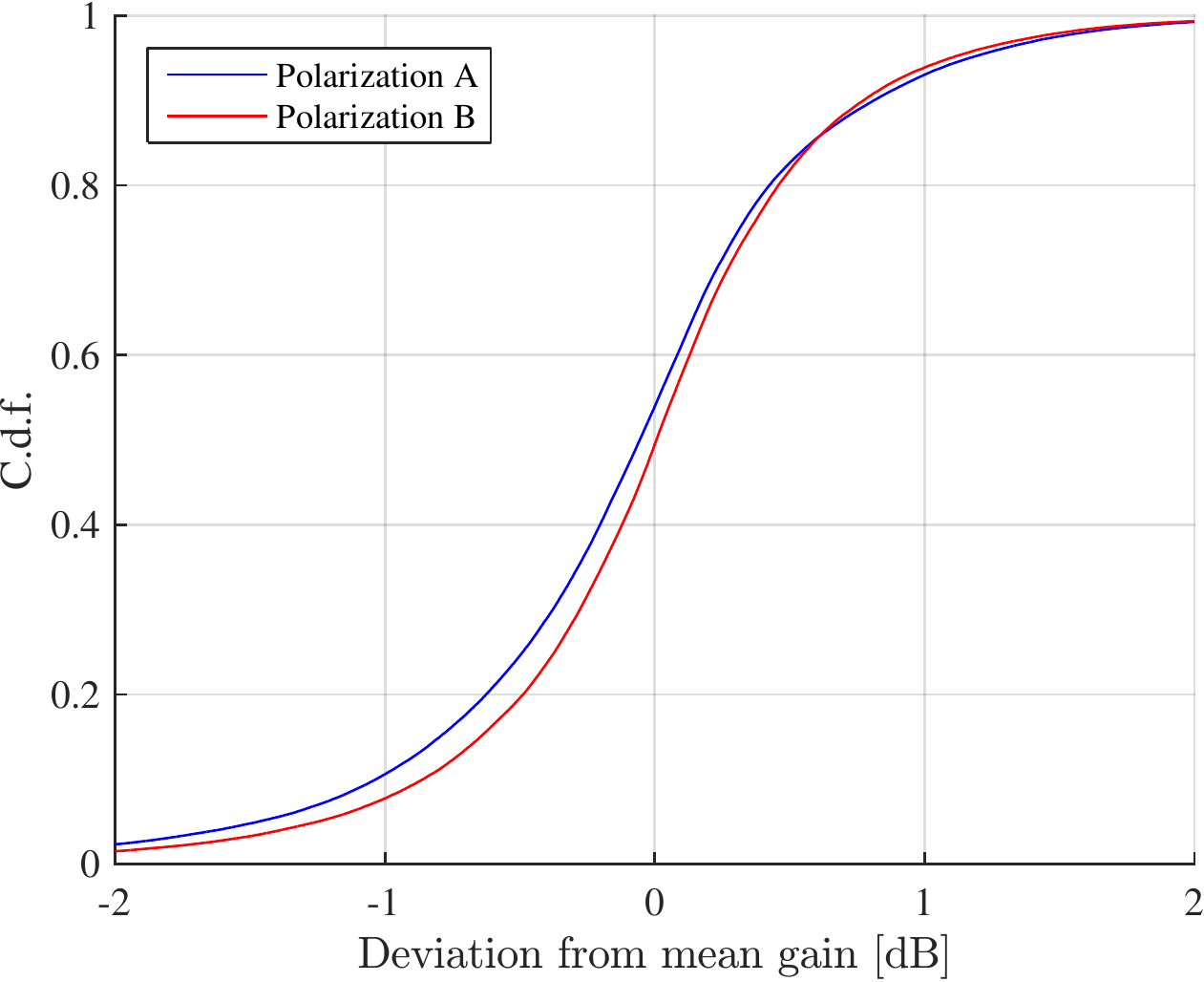}
    	\caption{Deviation in gain from the average subarray.}
    	\label{fig:cdfMeasSubarray}
	\end{subfigure}
	\caption{Realized gain of a subarray of the measured array.}
	\label{fig:realizedGainSubarray}
	\vspace{-0.5cm}
\end{figure}

\begin{figure}[ht]
	\centering
	\begin{subfigure}{0.5\textwidth}
		\centering
		\includegraphics[width=8cm, trim = 0.8cm 1cm 3cm 1cm, clip=True]{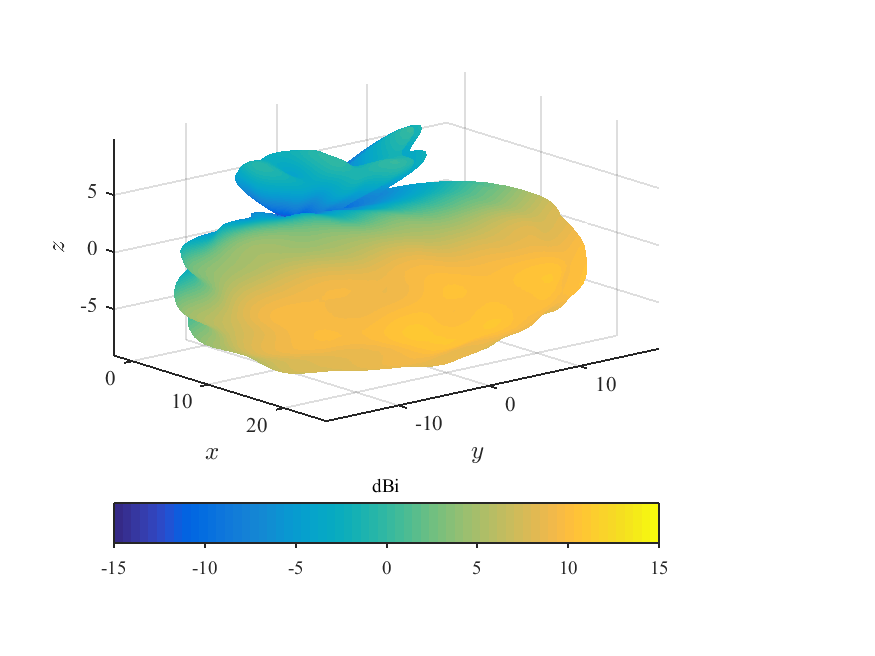}
    	\caption{Broad beam with the entire array.}
    	\label{fig:patternMeasArray}
	\end{subfigure}
	\begin{subfigure}{0.5\textwidth}
		\centering
		\includegraphics[width=8cm,trim = 0cm 0cm 0cm -0.5cm, clip=True]{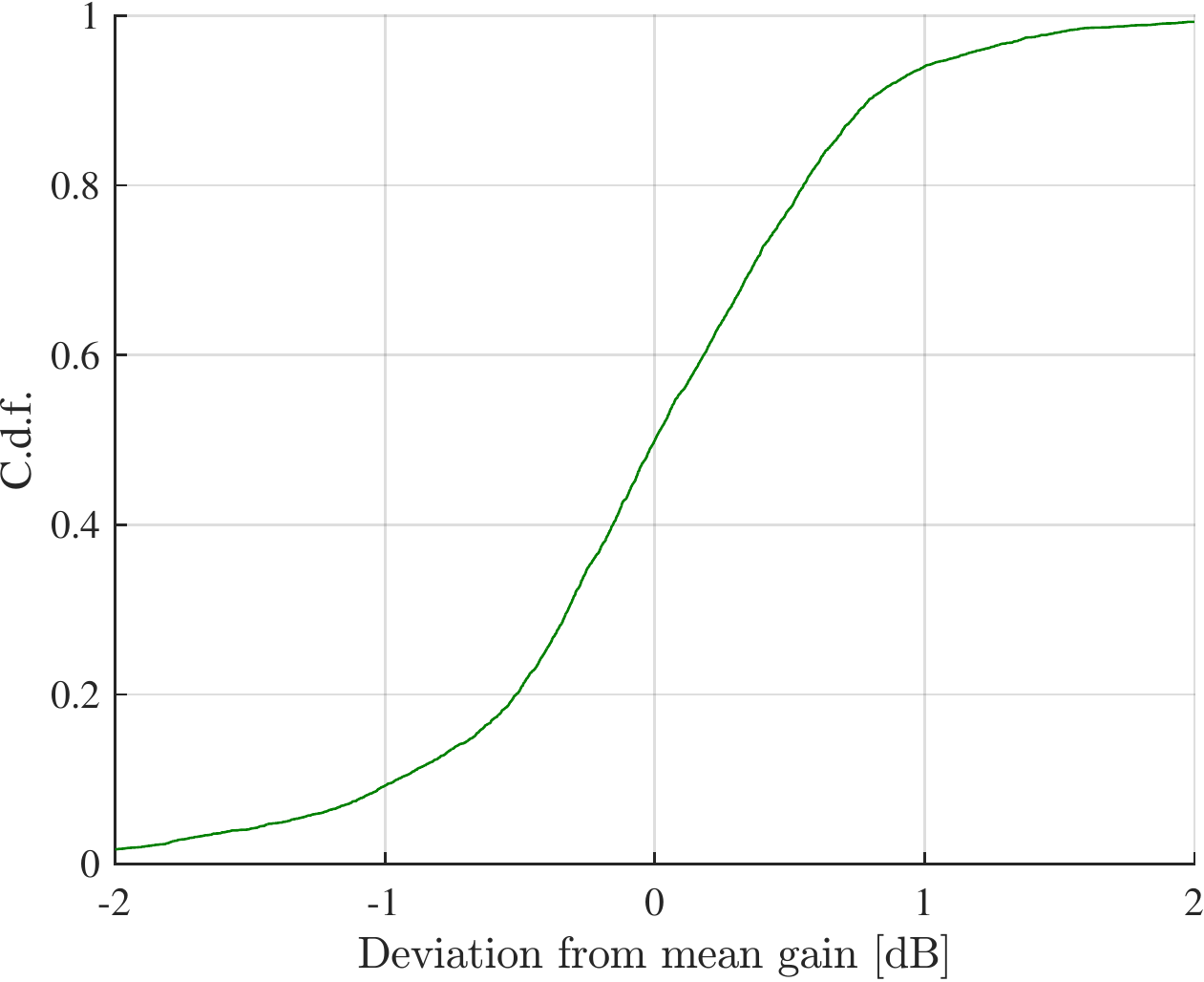}
    	\caption{Deviation in gain from the average subarray.}
    	\label{fig:cdfMeasArray}
	\end{subfigure}
	\caption{Realized gain for the entire measured array excited with $\epsilon$-complementary beamforming matrices of size $4\times 7$.}
	\label{fig:realizedGainArray}
	\vspace{-0.5cm}
\end{figure}

Here, we analyse how similar antenna elements are in real-world antenna array. The analysis is based on full-sphere measurements, provided by an antenna vendor, including both amplitude and phase for two orthogonal polarizations of an array with eight columns and four dual polarized subarrays per column (\ie, a $4\times 8$ rectangular array). The operation frequency is 3.8 GHz. The antenna elements are sub-arrays of dual-polarized elements having slanted $\pm45^{\circ}$ polarizations.

The analysis herein is based on \emph{realized gain}, meaning that both mismatch and ohmic losses are included in the measurement data~\cite{ieee2014standard}. The results of the analysis are presented in Fig.~\ref{fig:realizedGainSubarray}. It shall be noted that realized gain highlights antenna patterns and does not include benefits from efficient power utilization, which were demonstrated by the EIRP plots in Sec.~\ref{sec:simBroadBeamUla}. The benefit of this approach is that it gives us the ability to directly compare the beam shapes of a single element and the entire array, which will be useful in Appx.~B.
 
Based on the measurement data, we calculate total power patterns for each subarray. We then compute an \emph{average total power pattern} including all subarrays and both polarization ports (in total, 64 subarray ports), which is plotted in Fig.~\ref{fig:patternMeasSubarray}. Next, the difference, in dB, between the total power pattern for each of the 64 ports and the average power pattern was calculated for each $\brc{\phi, \theta}$ value within the service area and converted into two c.d.f.s (one per polarization) shown in Fig.~\ref{fig:cdfMeasSubarray}. As Fig.~\ref{fig:patternMeasSubarray} shows, the deviations for most subarrays and directions are very small---more than 80$\%$ of the data points fall within the range $[-1, 1]$ dB. Thus, we conclude that subarrays have quite similar patterns within the service area.

\section{Realistic Power-Efficient Broad Beams}
\label{sec:realisticBroadBeams}

Here we will, based on measurements of the same antenna as in Appx.~\ref{sec:realisticElements}, show that the concept described in Sec.~\ref{sec:broadBeamsArbitraryUla} actually works in real life. For that matter, we design a power-efficient broad beam for an array of size $4\times7$. The excitation matrices for both polarizations are obtained through the expansion of the $\epsilon$-complementary weights vectors~\eqref{eqn:weightsEpsCompl7} by means of~\eqref{eqn:weights2dVert} with expanders $\bx=[\j,1]^{\T}$ and $\by=[-1,-\j]^{\T}$. This gives us a pair of unimodular beamforming matrices.

Since the measured array has size $4\times 8$, while we need only $4\times7$ for our illustration, a column with all zeros was appended to each of our $\epsilon$-complementary beamforming matrices. Since these matrices yield a flat array factor, the obtained radiation pattern should be similar to the subarray pattern shown in Fig.~\ref{fig:patternMeasSubarray}. We plot the total power pattern of the power-efficient broad beam in Fig.~\ref{fig:patternMeasArray}, and, by comparing it with Fig.~\ref{fig:patternMeasSubarray}, we see that the patterns are indeed quite similar (subject to the ripple due to non-idealities of the measured array).

For the purpose of more detailed comparison of the patterns, a c.d.f. of the difference, in dB, between the two patterns within the service area is shown in Fig.~\ref{fig:cdfMeasArray}. Note that in this case we plot a single c.d.f., as the per-polarization patterns of the entire array do not look similar (\cf Fig.~\ref{fig:plotAacfAfExample}). Remarkably, the figure shows that the match between the patterns of the entire array and the average subarray is quite good. Moreover, the c.d.f. in Fig.~\ref{fig:cdfMeasArray} looks very similar to those in Fig.~\ref{fig:cdfMeasSubarray}. Based on these observations, one can conclude that, despite the assumption of identical power patterns and perfectly orthogonal polarization in~\eqref{eqn:eFieldsUla} are not fully met, it is still possible to create a power efficient broad beam with the proposed technique. The proposed method, thus, works well in practice for real-world antenna arrays and has already been used in numerous products of Ericsson.

\section*{Acknowledgements}
 The authors thank Fredrik Athley, Martin N. Johansson, Jonas Frid\'en, Henrik Asplund, George J\"{o}ngren, Johan Axn\"{a}s, Ulf Lindgren and Niklas Wernersson for fruitful discussions on the topic. We also thank the editor and anonymous reviewers for their helpful comments and suggestions.

\bibliographystyle{IEEEtran}
\bibliography{references}

\end{document}